\documentclass[secnumarabic,amssymb, nobibnotes, prl,twocolumn]{revtex4-1}
\usepackage{braket}
\usepackage{amsmath}
\usepackage{graphics}
\usepackage{graphicx}

\begin{document}

\title{Coherent Atom-Phonon Interaction through Mode Field Coupling in Hybrid Optomechanical Systems}%

\author{Michele Cotrufo}%
\email[Electronic address: ]{m.cotrufo@tue.nl}
\affiliation{Department of Applied Physics, Eindhoven University of Technology, 5600 MB Eindhoven, The Netherlands}
\author{Andrea Fiore}%
\affiliation{Department of Applied Physics, Eindhoven University of Technology, 5600 MB Eindhoven, The Netherlands}
\author{Ewold Verhagen}%
\affiliation{Center for Nanophotonics, FOM Institute AMOLF, Science Park 104, 1098 XG Amsterdam, The Netherlands}

\begin{abstract}
We propose a novel type of optomechanical coupling which enables a tripartite interaction between a quantum emitter, an optical mode and a macroscopic mechanical oscillator. The interaction uses a mechanism we term \textit{mode field coupling}: mechanical displacement modifies the spatial distribution of the optical mode field, which in turn modulates the atom-photon coupling rate. In properly designed multimode optomechanical systems, we can achieve situations in which mode field coupling is the only possible interaction pathway for the system. This enables, for example, swapping of a single excitation between emitter and phonon, creation of nonclassical states of motion and mechanical ground-state cooling in the bad-cavity regime. Importantly, the emitter-phonon coupling rate can be enhanced through an optical drive field, allowing active control of strong atom-phonon coupling for realistic experimental parameters.


\end{abstract}

\maketitle


Interfacing different quantum systems, such as atoms, photons, and phonons, is a key requirement for quantum information processing. The well-established framework of cavity quantum electrodynamics (CQED) interfaces photons --- ideal for communication --- to natural or artifical atoms (quantum emitters, QEs), whose strong nonlinearities enable quantum processing. Mechanical resonators have recently come to the forefront due to their large coherence times and their interaction with photons  in cavity optomechanical systems \cite{aspelmeyer2014cavity}. Moreover, the creation of nonclassical states in macroscopic mechanical systems is appealing for fundamental studies of quantum physics \cite{o2010quantum,lecocq2015resolving}. In these contexts, establishing an efficient and controllable interaction between phonons and QEs would be highly beneficial, as it would enable using the QE nonlinearity for the creation and manipulation of phononic quantum states \cite{pirkkalainen2015cavity}.

Different approaches have been proposed to realize such an interaction. First, a phonon can directly couple to a solid-state QE through mechanical strain \citep{golter2016,ramos2013,wilson2004}. Despite the large coupling rates obtainable in specific systems, this mechanism is difficult to engineer and to dynamically control. A second approach couples mechanical modes dispersively to an optical cavity, which in turn interacts with a QE \citep{restrepo2014,barzanjeh2011,chang2009,wang2010}. Tripartite entanglement  and atom-assisted optomechanical cooling are predicted in so-far elusive regimes when the optomechanical interaction is nonlinear at the quantum level \citep{restrepo2014} or when the emitter-field coupling rate approaches the emitter frequency \cite{chang2009}.
Additionally, QE-phonon interaction occurs in molecules and solids when the electronic and vibrational degrees of freedom are coupled, leading to inelastic scattering processes. Natural Raman transitions have been used to transfer a photon's quantum state to an optical phonon in diamond \cite{fisher2016,lee2012}, but the extremely high frequency and large dissipation limit general application for quantum processing.
\begin{figure}[th!] 	
  	\center
    \includegraphics[scale=0.55]{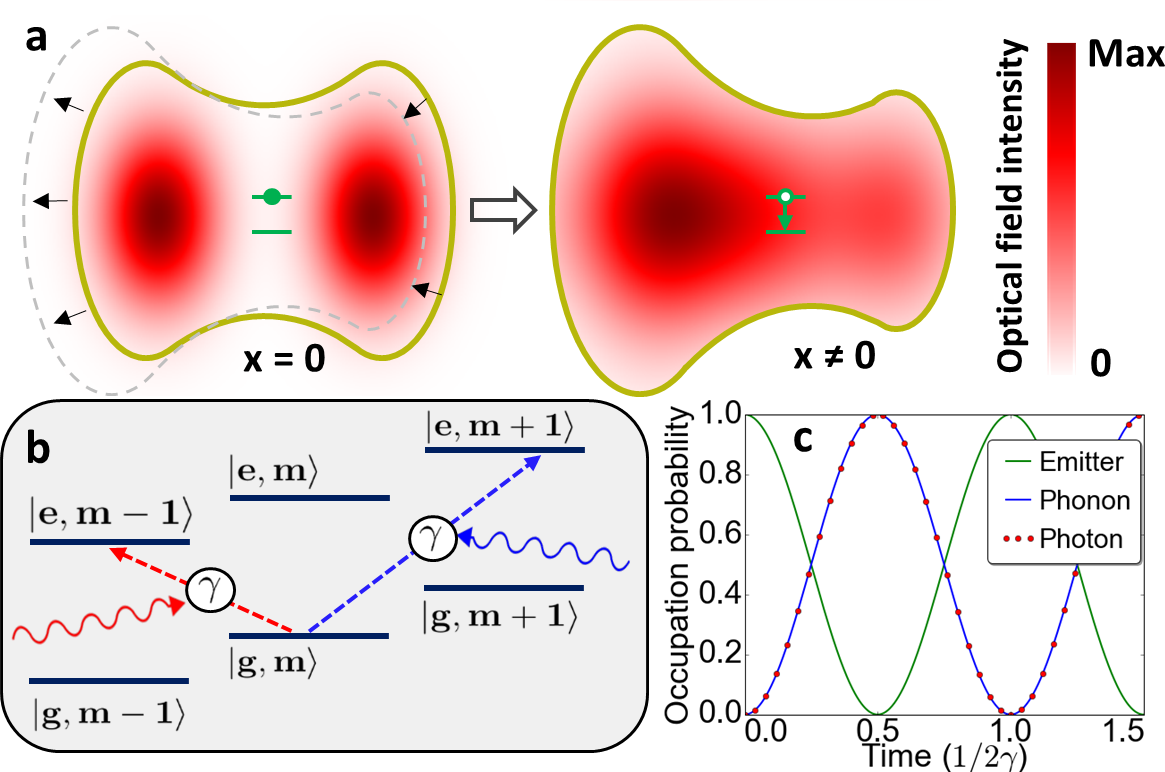}
    \caption{(color online). (a) Sketch of the proposed concept. An optical cavity (yellow solid line) defines an electric field (red pattern), which is initially (left part) zero at the QE position (green symbol). Upon displacement of the boundaries (right part), the field seen by the QE becomes non-zero and radiative transitions can occur.   (b) Phonon-non-conserving transitions achievable through MFC. A photon red-detuned (blue-detuned) by $\Omega_\mathrm{M}$ with respect to the cavity stimulates transitions  from $\ket{g,m}$ to $\ket{e,m-1}$ ($\ket{e,m+1}$). (c) Time  evolution dictated by the Hamiltonian in eq. \ref{eq_Hamitiltonian_VFC}, with the QE excited at $t=0$ and $\omega_\mathrm{c} = \omega_\mathrm{A} - \Omega_\mathrm{M}$.}\label{fig1_Intro}
\end{figure}

Here, we propose a novel optomechanical effect that provides an explicit, engineerable, and optically controllable interaction between a QE and a macroscopic mechanical oscillator. The interaction arises from a mechanically-induced modification of the spatial distribution of the optical field (fig. \ref{fig1_Intro}a), which in turn modulates the QE-photon coupling rate. We term this interaction \textit{mode field coupling} (MFC). We show that in simple multicavity optomechanical systems MFC is the only possible interaction for the system, enabling, \textit{e.g.}, QE-phonon excitation swapping and mechanical ground-state cooling in the bad-cavity regime. Importantly, the interaction strength can be controlled and enhanced by the optical field intensity, resulting in optically-controlled emitter-phonon coherent manipulation. This coupling, and the resulting Hamiltonian, share important traits with Raman-like processes in trapped ions \cite{leibfried2003}, which has proven powerful in controlling the motional state of single ions. MFC has however two distinct features: It involves large-mass macroscopic resonators, and its rate is nonetheless large enough to overcome the large decoherence typical of solid-state QEs.

\textit{Model.}
We consider a standard CQED setup, in which a two-level QE couples to an optical cavity mode through the Hamiltonian  $\hat{H} = \omega_\mathrm{A} \hat{\sigma}_z/2 + \omega_\mathrm{c} \hat{a}^\dag \hat{a} + g \left(\hat{a}\hat{\sigma}_+ + \hat{a}^\dag \hat{\sigma}_- \right)$, where $\omega_\mathrm{A}$ ($\omega_\mathrm{c}$) denotes the QE (optical mode) frequency, $\hat{a}$ is the photon annihilation operator, $\hat{\sigma}_{\pm,z}$ are Pauli operators describing the QE and $\hbar =1$. We initially neglect any loss, focusing on conservative interactions. The QE-field coupling rate $g = -\mathbf{d}\cdot \mathbf{\mathcal{E}}_0$ is determined by the emitter's transition dipole moment $\mathbf{d}$ and the electric field per photon $\mathbf{\mathcal{E}}_0$ of the optical mode at the emitter position. Next, we consider  a mechanical oscillator with frequency $\Omega_\mathrm{M}$ and phonon annihilation operator $\hat{b}$. In a standard dispersively coupled optomechanical system, the resonator's displacement  $\hat{x} = x_{\mathrm{zpf}} (\hat{b} + \hat{b}^\dag )$ affects the optical cavity frequency. This interaction is quantified by the coupling rate $g_0 = -(\partial \omega_\mathrm{c}/\partial x) x_{\mathrm{zpf}}$, where $x_{\mathrm{zpf}}$ is the zero-point motion amplitude. Here we consider a fundamentally different situation, in which the mechanical displacement induces a variation of the spatial distribution of the cavity field (fig. \ref{fig1_Intro}a), while $\omega_\mathrm{c}$ is negligibly affected. As a direct consequence the emitter-cavity coupling rate $g$ becomes dependent on mechanical position. Up to first order in $\hat{x}$, $g(\hat{x}) = g(0) + \gamma (\hat{b} + \hat{b}^\dag)$, where we defined the MFC coupling rate $\gamma =  (\partial g/\partial x)|_{x=0}  x_{\mathrm{zpf}}$.
Inserting this expression in the Hamiltonian $\hat{H}$ leads to the appearance of a tripartite interaction between the QE, the optical field, and the mechanical resonator.  In the specific case that at mechanical equilibrium the field at the emitter's position vanishes (fig. \ref{fig1_Intro}a), $g(0)=0$ and the only possible interaction channel for the system is the tripartite one. The interaction Hamiltonian reads
\begin{equation}
\label{eq_Hamitiltonian_VFC}
\hat{H}_\mathrm{int} =  \gamma  (\hat{b} + \hat{b}^\dag )\left(\hat{a}\hat{\sigma}_+ + \hat{a}^\dag \sigma_- \right),
\end{equation}
This Hamiltonian allows swapping the excitation between the three quantum systems under particular resonant conditions. 
For $\omega_\mathrm{c} \approx \omega_\mathrm{A} +   \Omega_\mathrm{M}$ ($\omega_\mathrm{c} \approx \omega_\mathrm{A} -   \Omega_\mathrm{M}$), the dominant term is $\hat{b}^\dag \hat{\sigma}_+ \hat{a} + h.c.$ ($\hat{b} \hat{\sigma}_+ \hat{a} + h.c.$), describing phonon and QE excitation upon annihilation of a photon (excitation of the QE due to photon and phonon annihilation) and the reverse process. Depending on the photon energy, therefore, the transitions $\ket{g,m} \leftrightarrow \ket{e,m\pm1}$ are realized (fig. \ref{fig1_Intro}b), where $e$ ($g$) denotes the QE excited (ground) state, and $m$ the phonon number. 
Figure \ref{fig1_Intro}c shows the lossless time evolution described by eq. \ref{eq_Hamitiltonian_VFC} for  $\omega_\mathrm{c} = \omega_\mathrm{A} - \Omega_\mathrm{M}$, with only the QE excited at $t=0$. The excitation oscillates, at a frequency $2\gamma$, between the QE and the state formed by one photon and one phonon.
Next, we consider pumping the cavity with a large coherent field to an average photon number $n_\mathrm{cav}$, writing the cavity field as $\hat{a} = \sqrt{n_{\mathrm{cav}}} + \delta \hat{a}$. Neglecting for now the fluctuations $\delta \hat{a}$ (valid for $n_{\mathrm{cav}}\gg1)$, the Hamiltonian reads
\begin{equation}
\label{eq_Hamitiltonian_VFC_Linearized}
\hat{H}_\mathrm{int} =  \gamma \sqrt{n_{\mathrm{cav}}} (\hat{b} + \hat{b}^\dag )\left(\hat{\sigma}_+ + \sigma_- \right),
\end{equation}
which describes a coherent QE-phonon interaction, with a coupling rate controlled by $n_{\mathrm{cav}}$. Thus, the optical intensity can enhance the QE-phonon coupling and, in particular, overcome system losses. 

\textit{Creating large field variation.} As the mode field is the solution of an eigenvalue problem \cite{joannopoulos2011photonic}, we look for a mechanical perturbation that induces strong changes of the eigenvector without affecting the eigenvalue. 
Such an effect is maximized for quasi-degenerate unperturbed eigenvalues, which can be obtained by coupling two or more cavities  such that hybridized modes (`supermodes') with well-defined symmetry are formed. 
Near a symmetry point, \textit{i.e.}, an anticrossing, an odd perturbation breaks symmetry. This localizes the supermodes in one of the cavities, resulting in a large variation of the local mode field.

\begin{figure*}
  	\centering
    \includegraphics[scale=0.59]{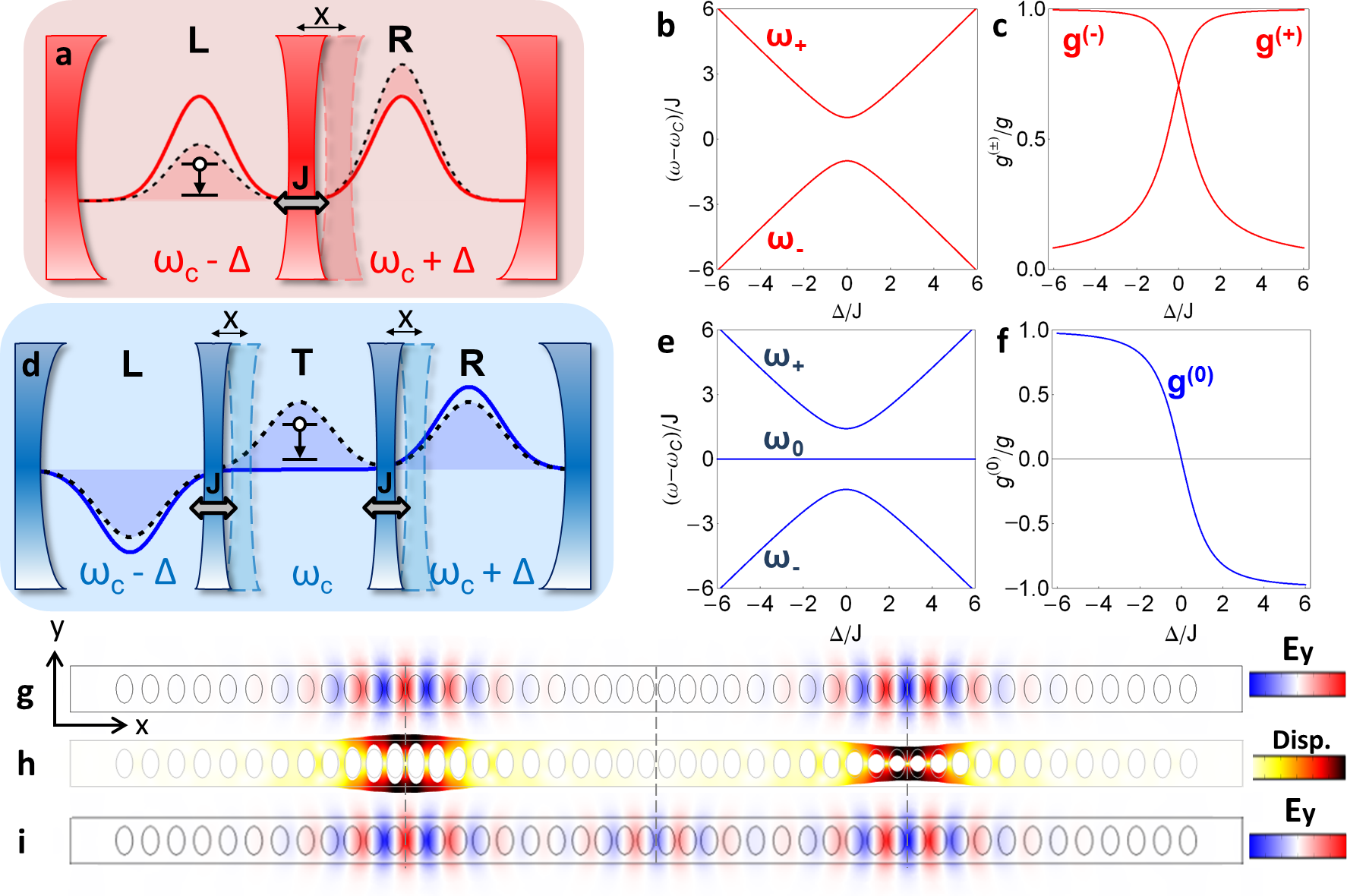}
    \caption{(color online). (a) Two identical cavities interact at a rate $J$ through a partly transparent movable mirror. The symmetric optical supermode is schematically depicted for the case of zero detuning (red line) and equal and opposite detuning $\pm \Delta$ on the cavities (dashed black line). (b) Supermode frequencies versus $\Delta/J$. (c) Coupling rates between a QE located in the left cavity (L) and the two supermodes (normalized to the coupling strength $g$ with the uncoupled mode in L) versus $\Delta/J$. (d) Three-cavity system.  The field of the supermode of interest, $\hat{a}_0$,  is shown as a blue line (black dashed line) for zero ($\pm \Delta$) detuning between the lateral cavities.  (e) Supermode frequencies in the three-cavity system. (f) Coupling rate between a QE located in cavity T and the supermode $\hat{a}_0$ (normalized to $g$). (g-i) Implementation based on three defect cavities in a photonic crystal nanobeam. Vertical dashed lines mark the cavity positions. (g) Electric field ($y$-component) of $\hat{a}_0$ at mechanical equilibrium. (h) Displacement pattern of the selected mechanical mode. (i) Expected electric field of $\hat{a}_0$ upon perturbation induced by the mechanical mode ($\Delta/J=0.5$). Additional details in \citep{SuppInfo}. }\label{fig2_2and3Cavities}
\end{figure*}
Indeed, we find an example of MFC (fig.  \ref{fig2_2and3Cavities}a) in two identical optical cavities coupled with rate $J$  (a \textit{membrane-in-the-middle} setup \citep{paraiso201,ludwig2012}). A mechanical displacement that induces opposite detuning $\pm \Delta$ to each cavity affects the spatial distribution of the supermode amplitudes, and thereby their coupling rate ($g^{(\pm)}$) with an emitter placed in one cavity (fig.  \ref{fig2_2and3Cavities}c). For $\Delta/J \ll 1$ the supermode frequencies are constant (fig. \ref{fig2_2and3Cavities}b), \textit{i.e.}, dispersive coupling is absent. The MFC coupling rate $\gamma$ scales as $J^{-1}$ (see supplemental information \cite{SuppInfo}): For weakly interacting cavities ($J\rightarrow 0$),  small deviations from the condition $\Delta=0$ quickly lead to localization of the supermodes into the individual cavities. In the two-cavity system, however, the tripartite MFC interaction competes with the Rabi emitter-photon interaction as $g^{(\pm)}(0) \neq 0$ (fig.  \ref{fig2_2and3Cavities}c).

This direct QE-photon interaction can be suppressed by introducing an additional optical cavity with identical frequency (fig. \ref{fig2_2and3Cavities}d). The middle cavity (T), which contains the QE, interacts with both lateral cavities with rate $J$, leading to the formation of three supermodes  $\hat{a}_+$, $\hat{a}_-$ and $\hat{a}_0$ \cite{johne2015,caselli2015}. The mode $\hat{a}_0$ has opposite fields in the lateral cavities and zero field in T (blue line in fig. \ref{fig2_2and3Cavities}d), and therefore does not interact with the QE. We now consider a mechanical mode that detunes only the frequencies of the lateral cavities by $\pm \Delta$, while leaving T unperturbed. This could be realized for example by rigidly connecting the two membranes. More generally, it can be obtained by dispersively coupling each optical cavity, at a rate $g_0$, to a mechanical oscillator \cite{SuppInfo}. If these three oscillators are coupled mechanically, one resulting mechanical supermode has equal and opposite dispersive interaction with the lateral cavities with a rate $\pm g_0/\sqrt{2}$, and zero interaction with T.
The frequency of $\hat{a}_0$ is unaltered by such detuning (fig. \ref{fig2_2and3Cavities}e), while its field in T assumes a finite value (fig. \ref{fig2_2and3Cavities}d, dashed line), which translates in a large modulation of the coupling rate $g^{(0)}$ between $\hat{a}_0$ and the QE  around the value $g^{(0)}=0$ (fig. \ref{fig2_2and3Cavities}f). 
Therefore, the interaction between the emitter, the mode $\hat{a}_0$ and the selected mechanical mode will be described by the Hamiltonian in eq. \ref{eq_Hamitiltonian_VFC}. 

Figure \ref{fig2_2and3Cavities}(g-h) shows an implementation of this model in a photonic crystal nanobeam. Cavities are defined by local variations of the periodicity, which results in  co-localized and dispersively coupled optical and mechanical resonances \citep{chan2011,burek2015diamond}. Three defect cavities are placed on the same nanobeam, leading to both optical and mechanical hybridization. Inter-cavity separation controls the optical interaction rate $J$. The electric field of $\hat{a}_0$ (fig.  \ref{fig2_2and3Cavities}g) is zero in the central cavity when the mechanical mode is at rest. Figure \ref{fig2_2and3Cavities}h shows the mechanical mode that provides the required detuning on the lateral cavities. Upon mechanically-induced detuning of the lateral cavities, the mode $\hat{a}_0$ acquires a finite electric field in the central cavity (fig. \ref{fig2_2and3Cavities}i).

\begin{figure}[th!] 	
  	\center
    \includegraphics[scale=0.55]{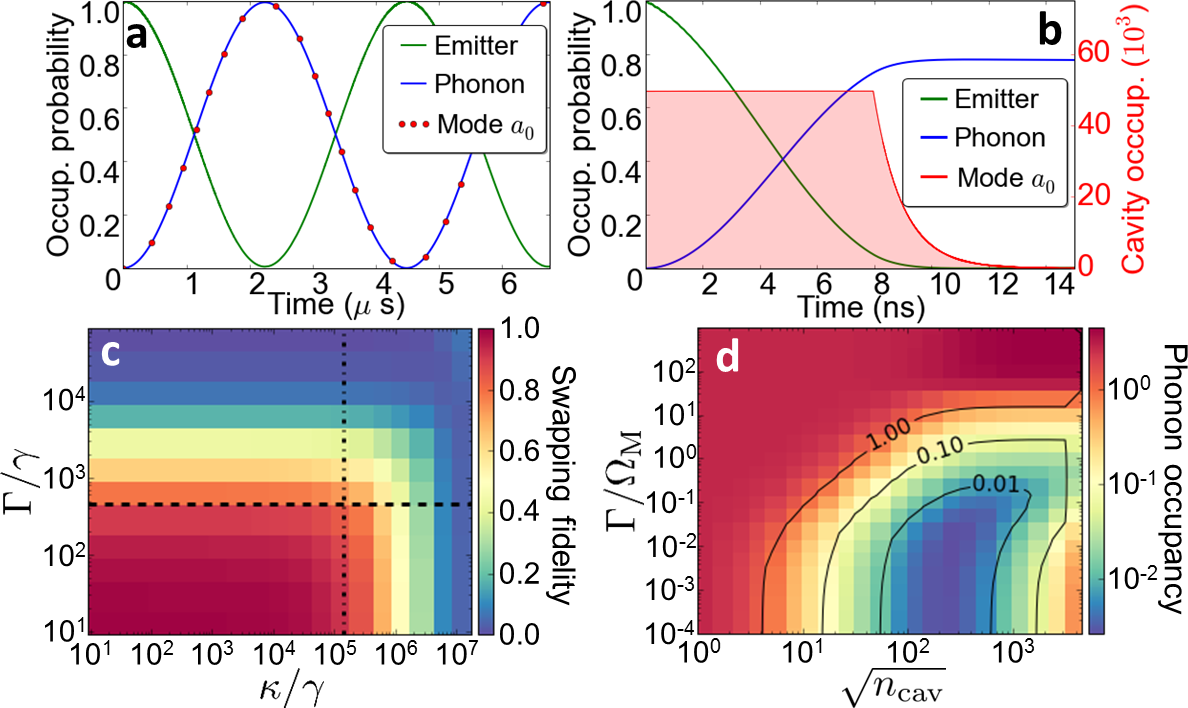}
    \caption{(color online). Numerical calculations based on the three-cavity Hamiltonian (eq.\ref{eq_Full3Cavities_Hamiltonian}, parameters in text). (a) Lossless evolution starting from an excited QE without optical pumping. (b) As in (a) but assuming losses for the QE  and optical cavity. The mode $\hat{a}_0$ is continuosly pumped until t $\approx$ 8 ns, swapping the excitation from emitter to resonator and creating a nonclassical motional state. (c) Fidelity of the QE-phonon swapping versus  $\Gamma$ and $\kappa$ for $n_{\mathrm{cav}}=5 \cdot 10^4$. The horizontal dashed line indicates the condition $\Gamma=\gamma \sqrt{n_{\mathrm{cav}}}$, while the vertical dashed-dotted line indicates $\kappa =2g_0 J \sqrt{n_{\mathrm{cav}}}/g$. (d) Cooling through MFC. Steady-state mean phonon number versus $\Gamma$ and $n_{\mathrm{cav}}$ for continuous optical pumping and $\kappa=10\Omega_\mathrm{M}$.} \label{fig3_NumericalCalculations1}
\end{figure}

\textit{Full model and numerical calculations.}
We now analyse the three-cavity system in detail, and show that it behaves as predicted by the Hamiltonian in eq. \ref{eq_Hamitiltonian_VFC}. For simplicity, we consider only one of  the hybridized mechanical supermodes, described by the operator $\hat{b}$, frequency $\Omega_\mathrm{M}$ and dispersively coupled to the lateral cavities at a rate $\pm g_0/\sqrt{2}$. The validity of this approach is justified in \cite{SuppInfo}. In a frame rotating at $\omega_\mathrm{c}$, the Hamiltonian is
\begin{align}
\label{eq_Full3Cavities_Hamiltonian}
\begin{split}
\hat{H} &= -\hat{\Delta} \hat{a}^\dag_L  \hat{a}_\mathrm{L} + \hat{\Delta}  \hat{a}^\dag_R  \hat{a}_\mathrm{R} + \Omega_\mathrm{M} \hat{b}^\dag \hat{b} + \frac{\omega_\mathrm{A}-\omega_\mathrm{c}}{2}\hat{\sigma}_z + \\ & + J \left[ \hat{a}^\dag_T (\hat{a}_\mathrm{L} +\hat{a}_\mathrm{R}) + h.c. \right] + g\left(  \hat{a}_\mathrm{T} \hat{\sigma}_+ + h.c. \right),
\end{split}
\end{align}
where we defined $\hat{\Delta} = g_0/\sqrt{2}( \hat{b} + \hat{b}^\dag )$. The first two terms describe the mechanically-induced detuning on the lateral cavities. The  second row describes optical mode coupling and the Rabi interaction between emitter and cavity T. 
Assuming a quasi-static approximation \citep{ludwig2012} for the mechanical motion (valid for $J\gg \Omega_\mathrm{M}$), we can treat $\hat{\Delta}$ quasi-statically and diagonalize the optical part of the Hamiltonian by introducing three optical supermodes, $\hat{a}_\pm$ and $\hat{a}_0$.
Up to the first order in $\hat{\Delta}/J$, we obtain an interaction Hamiltonian
\begin{align}
\label{eq_Full3Cavities_Hamiltonian_CoupledModes_FirstOrder}
\begin{split}
&\hat{H}_\mathrm{int} = \gamma (\hat{b} + \hat{b}^\dag )\left(\hat{a}_0\hat{\sigma}_+ + h.c. \right) + g'\left[\left(\hat{a}_+ - \hat{a}_-  \right)\hat{\sigma}_+ + h.c.\right].
\end{split}
\end{align}
The first term of eq. \ref{eq_Full3Cavities_Hamiltonian_CoupledModes_FirstOrder} shows the tripartite interaction explicitly, with $\gamma=g g_0/(2J)$. The last term describes a Rabi interaction between the emitter and the supermodes $\hat{a}_\pm$ with coupling rate $g'=g/\sqrt{2}$. A pure tripartite interaction can therefore be obtained for large supermode separation ($J\gg \Omega_\mathrm{M}$) and tripartite resonance ($\omega_\mathrm{A} \approx \omega_\mathrm{c} \pm \Omega_\mathrm{M}$).  Additionally, to let the emitter interact with the supermodes (and not the uncoupled modes), we require $J\gg g$. To verify that the predicted coherent emitter-phonon interaction occurs in a realistic scenario, we numerically solve \cite{johansson2012qutip} the master equation derived from the full Hamiltonian in eq. \ref{eq_Full3Cavities_Hamiltonian}. Out of the many possible systems, we consider the structure of fig. \ref{fig2_2and3Cavities}(g-h) made in diamond with a nitrogen vacancy (NV) center as emitter. The simulated parameters for this system are \{$\omega_\mathrm{c}$, $\Omega_\mathrm{M}$, $g$, $g_0$\} = $2\pi\cdot$\{4.7$\cdot 10^5$, 14, 20, 0.004\} GHz. We consider the case $\omega_\mathrm{A} = \omega_\mathrm{c} +\Omega_\mathrm{M}$ and $J=18g$, corresponding to one period cavity separation \citep{SuppInfo}. The unitary evolution of the system starting with the emitter excited (fig. \ref{fig3_NumericalCalculations1}a) agrees perfectly with that of the MFC Hamiltonian (fig. \ref{fig1_Intro}c) and verifies the predicted QE-phonon oscillation period $\pi/\gamma$ = 4.5 $\mu$s. This confirms that for realistic choices of parameters purely tripartite interaction is obtained in the three-cavity system. In order to overcome losses, unavoidable in an experimental setting, the coupling rate can be enhanced by selectively pumping the supermode $\hat{a}_0$ \citep{SuppInfo}. Figure \ref{fig3_NumericalCalculations1}b shows the evolution of the same system as in fig. \ref{fig3_NumericalCalculations1}a, now with dissipations introduced through Lindblad operators \citep{SuppInfo}, with a cavity decay rate $\kappa/2\pi$ = 1 GHz and a conservative emitter decay rate $\Gamma/2\pi$ = 0.05 GHz (measured for NV centers in photonic crystal structures \cite{lee2014}). In fact, much smaller radiative decay rates are in principle expected here \citep{SuppInfo}.  The mode $\hat{a}_0$ is continuously  pumped to a steady-state population of  $n_{\mathrm{cav}}$ = $5\cdot 10^4$, shown to be experimentally feasible in diamond \cite{burek2015diamond}. We note that the emitter is not directly affected by the large optical intensity as the field is zero at the emitter position. At $t = 0$ the QE is excited and interacts with the mechanical mode with a coupling rate $\gamma \sqrt{n_{\mathrm{cav}}}$. At $t \approx \pi/2\gamma \sqrt{n_{\mathrm{cav}}}$ the pump is switched off, suppressing the interaction and leaving the system in a long-lived nonclassical state with phonon population $n_b\approx$ 0.8. 
The swapping fidelity can be made arbitrarily close to one (fig. \ref{fig3_NumericalCalculations1}c) by reducing the QE decay rate (so that $\Gamma \ll \gamma \sqrt{n_{\mathrm{cav}}}$) and the optical losses $\kappa$ of the supermodes $\hat{a}_\pm$, which introduce additional decay for the QE due to the finite optical linewidth. This decay is negligible when $\kappa \ll 2 g_0 J  \sqrt{n_{\mathrm{cav}}} / g $ (vertical dashed-dotted line in fig. \ref{fig3_NumericalCalculations1}c) \citep{SuppInfo}. The influence of  $n_{\mathrm{cav}}$ and QE dephasing  on the swapping fidelity is discussed in the supplemental information \citep{SuppInfo}.

The proposed QE-phonon interaction can also be used to cool the mechanical resonator to its ground state. The cooling cycle is triggered by a red-detuned cavity photon which excites the QE upon annihilation of a phonon. The excitation is subsequently dissipated through the QE decay. Differently from standard optomechanical cooling \citep{aspelmeyer2014cavity}, this mechanism can achieve ground state cooling in the bad cavity regime ($\kappa \gg \Omega_\mathrm{M}$), while the sideband resolved regime is required only for the QE ($\Gamma < \Omega_\mathrm{M}$). Figure \ref{fig3_NumericalCalculations1}d shows the steady-state phonon population in the three-cavity system as a function of $\Gamma$ and $n_\mathrm{cav}$, for $\kappa=10\Omega_\mathrm{M}$ and for finite mechanical losses ($\Gamma_\mathrm{M}/2\pi=50$ kHz) and thermal phonon occupation ($n_\mathrm{th}=4$). As expected, ground-state cooling is possible for $\Gamma/\Omega_\mathrm{M} \lesssim  1$. Phonon population lower than 0.1 can be achieved with $n_\mathrm{cav} \lesssim 10^3$ and realistic QE decay rates. The phonon population increase for large $n_\mathrm{cav}$ is attributed to the onset of ultra-strong coupling, as $\gamma \sqrt{n_\mathrm{cav}}$ approaches $\Omega_\mathrm{M}$. For small $\Gamma$, the QE total decay rate is dominated by the additional emission into the supermodes $\hat{a}_\pm$ (which read $\Gamma_{\pm}=g^2\kappa/4 J^2 \approx 0.11\cdot 2\pi$ GHz \citep{SuppInfo}), which explains the saturation of the phonon population for $\Gamma/\Omega_\mathrm{M} < 10^{-2}$.

In conclusion, we have introduced a new kind of emitter-photon-phonon interaction in hybrid-optomechanical systems, based on mechanically-induced variation of the electric field spatial pattern. The coupling rate can be particularly strong in multicavity systems with small coupling rate, as it scales inversely with the coupling rate $J$. For large optical drives, this mechanism leads to an emitter-phonon coherent interaction whose coupling strength is controlled by the optical intensity. Emitter-phonon excitation swapping and mechanical ground-state cooling are possible with feasible experimental parameters. The proposed interaction strength is much larger than effects obtainable in single-mode systems, which require the ultra-strong coupling regime ($g\approx \omega_\mathrm{c}$) to have comparable rates \citep{barzanjeh2011,chang2009,wang2010,SuppInfo}. Differently from strain-based methods \citep{golter2016,ramos2013,wilson2004}, the proposed coupling mechanism is not limited to a specific choice of the solid-state emitter and material system, and it could even be applied to atoms trapped near a mechanical resonator \cite{goban2014atom,thompson2013coupling,hammerer2009strong}. Moreover, it provides strong quantum nonlinearity without requiring the single-photon strong optomechanical coupling regime ($g_0\gg\kappa$). In perspective, the optically controlled coherent emitter-interaction introduced here paves the way for, \textit{e.g.}, control of spontaneous phonon emission, creation of nonclassical states of motion and phonon lasing.
 
%


\begin{acknowledgments}
The authors acknowledge L. Midolo for first pointing out the possibility of displacement-induced field variations in optomechanical cavities. This work is part of the research programme of the Foundation for Fundamental Research on Matter (FOM), which is financially supported by the Netherlands Organisation for Scientific Research (NWO). E.V. gratefully acknowledges an NWO-Vidi grant for financial support.
\end{acknowledgments}
%
\newpage

\renewcommand{\theequation}{S\arabic{equation}}
\renewcommand{\thefigure}{S\arabic{figure}}
\onecolumngrid
\begin{center}
\Large
{\textbf{Supplementary Material}}
\end{center}
\large
\section{Density matrix calculations, numerical setup}
\label{sec_MasterEQ}
The non-Hermitian evolution of the different systems considered in this work has been calculated with a Master equation approach, in which the different dissipative channels are described by proper Lindblad terms. For a given Hamiltonian $\hat{H}$, we calculate the temporal evolution of the density matrix $\rho$ through the equation
\begin{align}
\label{eq_MasterEquation}
\begin{split}
\dot{\hat{\rho}} &= - \frac{i}{\hbar}\left[\hat{H},\hat{\rho} \right ] + \sum_{i} \frac{\kappa_i}{2} \left(2\hat{a}_i \hat{\rho } \hat{a}_i^\dag - \left\{\hat{a}_i^\dag \hat{a}_i, \hat{\rho}\right\} \right) + \frac{\Gamma_\mathrm{M}(n_\mathrm{th}+1)}{2} \left(2\hat{b} \hat{\rho } \hat{b}^\dag - \left\{\hat{b}^\dag \hat{b}, \hat{\rho}\right\} \right) +\\&+\frac{\Gamma_\mathrm{M} n_\mathrm{th}}{2} \left(2\hat{b}^\dag \hat{\rho } \hat{b} - \left\{\hat{b} \hat{b}^\dag, \hat{\rho}\right\} \right) + \frac{\Gamma}{2} \left(2\hat{\sigma}_- \hat{\rho } \hat{\sigma}_+ - \left\{\hat{\sigma}_+ \sigma_-, \hat{\rho}\right\} \right) + \frac{\Gamma^*}{2} \left(\hat{\sigma}_z \hat{\rho } \hat{\sigma}_z -\hat{\rho} \right)
\end{split}
\end{align}
where the sum refers to all the optical cavities considered, $\kappa_i$ is decay rate of the i-th cavity, $\Gamma_\mathrm{M}$ is the decay rate of the mechanical mode, $n_\mathrm{th}$ is the average phonon number of the external bath, and $\Gamma$ and $\Gamma^*$ are the decay rate and pure dephasing rate of the emitter, respectively. The brackets [,] and \{,\} indicate commutation and anti-commutation of the operators, respectively.
The master equation has been solved numerically with the opensource Python framework QuTIP \cite{SI_johansson2012qutip}. For the numerical calculations, the dimensions of the Fock spaces of the optical cavities and mechanical resonator need to be truncated. For the calculations of the emitter-phonon swapping, the Fock spaces of the mechanical oscillator and optical cavities all have dimensions of 2. We verified that no appreciable numerical differences arise for larger Fock spaces dimensions. For the cooling calculations, the Fock space of the mechanical resonator has a dimension of 15, to ensure that a thermal state with $n_\mathrm{th}=4$ can be properly described.

\section{Derivation of the Mode Field Coupling for the two-cavity system}
\begin{figure}[th!] 	
  	\center
    \includegraphics[scale=0.7]{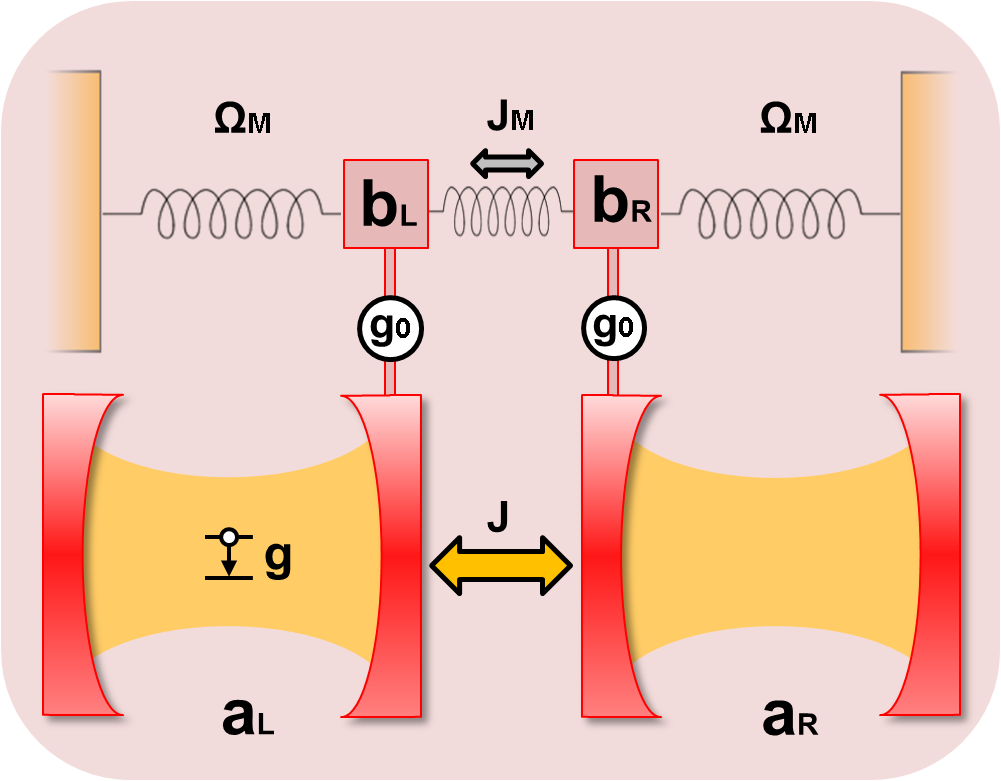}
    \caption{Schematic of the two-cavity system. Two identical optical cavities (each denoted by a couple of red mirrors) interact with each other at a rate $J$. Each cavity is dispersively coupled to a separate mechanical resonator with an optomechanical coupling rate $g_0$. The resonators are identical and have frequency $\Omega_\mathrm{M}$. The two mechanical resonators are additionally coupled to each other at a rate $J_\mathrm{M}$. An emitter is placed in the left optical cavity and interacts with one of its optical modes with a coupling rate $g$.}\label{figSuppInfo_Model_2_Cavities.png}
\end{figure}

The general model describing mode field coupling in a two-cavity system is schematically depicted in fig. \ref{figSuppInfo_Model_2_Cavities.png}. We consider two identical optomechanical systems, denoted left (L) and right (R), each composed of an optical cavity with frequency $\omega_\mathrm{c}$ and a mechanical resonator with frequency $\Omega_\mathrm{M}$. In each system, the cavity and the resonator are dispersively coupled at a rate $g_0$. We describe the fields in the two optical cavities with the annihilation operators $\hat{a}_\mathrm{L}$ and $\hat{a}_\mathrm{R}$, and the two mechanical resonators with annihilation operators $\hat{b}_\mathrm{L}$ and $\hat{b}_\mathrm{R}$. The two optical cavities are coupled with a rate $J$, while the two mechanical resonators are coupled with a rate $J_\mathrm{M}$. Finally, a two-level emitter is placed in the left cavity, and interacts with the field $\hat{a}_\mathrm{L}$ with a coupling rate $g$.

The full Hamiltonian reads
\begin{equation}
\begin{split}
\hat{H} = &\left[\omega_\mathrm{c} - g_0 \left(\hat{b}_\mathrm{L}^\dag +\hat{b}_\mathrm{L} \right) \right] \hat{a}_\mathrm{L}^\dag \hat{a}_\mathrm{L} + \left[\omega_\mathrm{c} - g_0 \left(\hat{b}_\mathrm{R}^\dag +\hat{b}_\mathrm{R} \right) \right] \hat{a}_\mathrm{R}^\dag \hat{a}_\mathrm{R} + \Omega_\mathrm{M} \left(\hat{b}_\mathrm{L}^\dag \hat{b}_\mathrm{L} + \hat{b}_\mathrm{R}^\dag \hat{b}_\mathrm{R}\right) + \frac{\omega_\mathrm{A}}{2}\hat{\sigma}_z +\\
&+J \left(\hat{a}^\dag_R \hat{a}_\mathrm{L} + h.c.\right) + J_\mathrm{M} \left(\hat{b}^\dag_R \hat{b}_\mathrm{L} + h.c.\right)+ g\left(\hat{\sigma}_+ \hat{a}_\mathrm{L} + h.c. \right).
\end{split}
\end{equation}
We now introduce mechanical supermodes $\hat{b}_{\pm} = \frac{1}{\sqrt{2}} \left(\hat{b}_\mathrm{L} \pm \hat{b}_\mathrm{R} \right)$. Substituting these in the Hamiltonian, we obtain
\begin{equation}
\begin{split}
\hat{H} =& \left[\omega_\mathrm{c} - \frac{g_0}{\sqrt{2}} \left(\hat{x}_+ + \hat{x}_- \right) \right] \hat{a}_\mathrm{L}^\dag \hat{a}_\mathrm{L} + \left[\omega_\mathrm{c} - \frac{g_0}{\sqrt{2}} \left(\hat{x}_+ - \hat{x}_- \right) \right] \hat{a}_\mathrm{R}^\dag \hat{a}_\mathrm{R} + \\ &+ \left(\Omega_\mathrm{M}+J_\mathrm{M} \right) \hat{b}_+^\dag \hat{b}_+ + \left(\Omega_\mathrm{M}-J_\mathrm{M}\right) \hat{b}_-^\dag \hat{b}_-  + \frac{\omega_\mathrm{A}}{2}\hat{\sigma}_z +\\
&+J \left(\hat{a}^\dag_R \hat{a}_\mathrm{L} + h.c.\right) + g\left(\hat{\sigma}_+ \hat{a}_\mathrm{L} + h.c. \right),
\end{split}
\end{equation}
where we have defined the dimensionless position operators of the mechanical supermodes $\hat{x}_{\pm} = \hat{b}_{\pm}^\dag +\hat{b}_{\pm}$. The mechanical mode $\hat{b}_-$ interacts dispersively with the modes $\hat{a}_\mathrm{R}$ and $\hat{a}_\mathrm{L}$ with equal and opposite coupling rate $\pm g_0/\sqrt{2}$. In the following, we will neglect the presence of the other mechanical supermode, $\hat{b}_+$. For simplicity, we replace $\hat{b}_- \rightarrow \hat{b}$ and $\left(\Omega_\mathrm{M} - J_\mathrm{M} \right) \rightarrow \Omega_\mathrm{M}$. We moreover define the operator $\hat{\Delta} = \dfrac{g_0}{\sqrt{2}} \left( \hat{b}^\dag +\hat{b}\right)$. The new Hamiltonian reads
\begin{equation}
\begin{split}
\hat{H} =& -\hat{\Delta} \hat{a}_\mathrm{L}^\dag \hat{a}_\mathrm{L} +\hat{\Delta}  \hat{a}_\mathrm{R}^\dag \hat{a}_\mathrm{R} + \Omega_\mathrm{M} \hat{b}^\dag \hat{b}  + \frac{\omega_\mathrm{A}-\omega_\mathrm{c}}{2}\hat{\sigma}_z +\\
&+J \left(\hat{a}^\dag_R \hat{a}_\mathrm{L} + h.c.\right) + g\left(\hat{\sigma}_+ \hat{a}_\mathrm{L} + h.c. \right),
\end{split}
\end{equation}
where we also performed a unitary transformation $\hat{H} \rightarrow  \hat{U}(t) \hat{H} \hat{U}^\dag(t) - i \hat{U}(t) \dfrac{\partial \hat{U}(t)^\dag}{\partial t}$, with $\hat{U}(t) = \exp\left[-i\omega_\mathrm{c} t \left(\hat{a}_\mathrm{L}^\dag \hat{a}_\mathrm{L}+ +\hat{a}_\mathrm{R}^\dag \hat{a}_\mathrm{R} + \hat{\sigma}_+ \hat{\sigma}_- \right) \right]$. Assuming a quasi-static approximation for $\hat{\Delta}$, valid in the limit $J\gg\Omega_\mathrm{M}$, we can diagonalize the optical part of the Hamiltonian \citep{SI_ludwig2012}. The optical supermodes are defined by
\begin{equation}
\begin{split}
\label{eq_OpticalNormalModes}
\hat{a}_\mathrm{+} = \alpha \hat{a}_\mathrm{L} + \beta \hat{a}_\mathrm{R}, \;
\hat{a}_\mathrm{-} = \beta \hat{a}_\mathrm{L} - \alpha\hat{a}_\mathrm{R},
\end{split}
\end{equation}
where $\alpha$ and $\beta$ are operators defined by
\begin{equation}
\label{eq_OpticalNormalModesAlphaBeta}
\alpha = \frac{\sqrt{\hat{\Delta}^2 + J^2} + \hat{\Delta}}{\sqrt{\left(\sqrt{\hat{\Delta}^2 + J^2} + \hat{\Delta}\right)^2 + J^2}}, \;\;
\beta = \frac{J}{\sqrt{\left(\sqrt{\hat{\Delta}^2 + J^2} + \hat{\Delta}\right)^2 + J^2}}.
\end{equation}
In the supermode basis, the Hamiltonian is
\begin{align}
\label{eq_Hamiltonian2Cavites}
\begin{split}
\hat{H} =& \left(\omega_\mathrm{c} + \sqrt{J^2+\hat{\Delta}^2}\right) \hat{a}^\dag_+ \hat{a}_\mathrm{+} + \left(\omega_\mathrm{c} - \sqrt{J^2+\hat{\Delta}^2}\right) \hat{a}^\dag_- \hat{a}_\mathrm{-} +  \omega_\mathrm{A} \frac{\hat{\sigma}_z}{2} + \Omega_\mathrm{M} \hat{b}^\dag \hat{b} +\\
 &	+ g \beta \left( \hat{\sigma}_+ \hat{a}_\mathrm{-} + h.c. \right).
\end{split}
\end{align}
Up to the first order in $\hat{\Delta}/J$, $\alpha$ and $\beta$ read
\begin{equation}
\alpha = \frac{1}{\sqrt{2}}\left(1  + \frac{\hat{\Delta}}{2J}\right) +  \mathcal{O}\left(\frac{\hat{\Delta}}{J}\right)^2, \;\;
\beta = \frac{1}{\sqrt{2}}\left(1 - \frac{\hat{\Delta}}{2J}\right)  +  \mathcal{O}\left(\frac{\hat{\Delta}}{J}\right)^2.
\end{equation}
By inserting these expansions in the Hamiltonian in eq.~\ref{eq_Hamiltonian2Cavites}, and expressing again $\hat{\Delta}$ as a function of $\hat{b}$ and $\hat{b}^\dag$, we get
\begin{align}
\label{eq_Hamiltonian2Cavites_Linearized}
\begin{split}
\hat{H} =& \left(\omega_\mathrm{c} + \sqrt{J^2+\hat{\Delta}^2}\right) \hat{a}^\dag_+ \hat{a}_\mathrm{+} + \left(\omega_\mathrm{c} - \sqrt{J^2+\hat{\Delta}^2}\right) \hat{a}^\dag_- \hat{a}_\mathrm{-} +  \omega_\mathrm{A} \frac{\hat{\sigma}_z}{2} + \Omega_\mathrm{M} \hat{b}^\dag \hat{b} +\\
 &+\frac{g}{\sqrt{2}} \left( \hat{\sigma}_+ \hat{a}_\mathrm{+} + h.c. \right) +\frac{g}{\sqrt{2}}  \left( \hat{\sigma}_+ \hat{a}_\mathrm{-} + h.c. \right) + \\
 &+\frac{g g_0}{4J} \left( \hat{b}^\dag +\hat{b}\right) \left( \hat{\sigma}_+ \hat{a}_\mathrm{+} + h.c. \right) -\frac{g g_0}{4J}  \left( \hat{b}^\dag +\hat{b}\right)\left( \hat{\sigma}_+ \hat{a}_\mathrm{-} + h.c. \right).
\end{split}
\end{align}
The second row describes the Rabi interaction of the emitter with the two optical supermodes, with a coupling rate $g/\sqrt{2}$. As mentioned in the main text, in the two-cavity system this interaction is unavoidable and is due to the fact that both optical supermodes have nonzero field in both cavities for all finite detunings. The third row describes the tripartite interaction between the emitter, the phonon and the optical supermodes.

\section{Derivation of the Mode Field Coupling for the three-cavity system}
\begin{figure}[th!] 	
  	\center
    \includegraphics[scale=0.7]{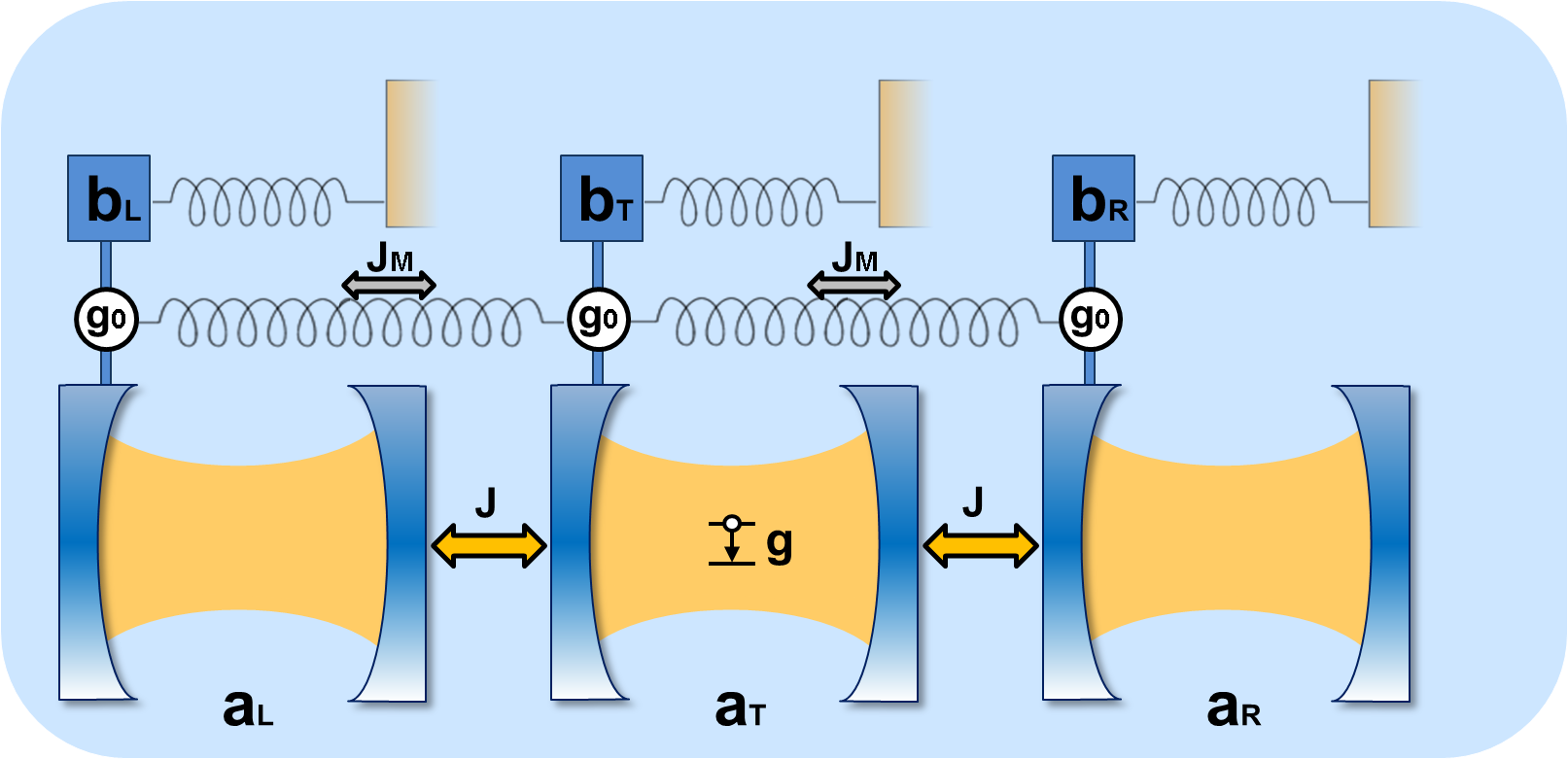}
    \caption{Schematic of the three-cavity system. Three identical optical cavities (each denoted by a couple of blue mirrors) are arranged such that the central one interacts with the two lateral ones with a rate $J$. Each cavity is dispersively coupled to a separate resonator with an optomechanical coupling rate $g_0$. The resonators are identical and have frequency $\Omega_\mathrm{M}$. The central mechanical resonator is additionally coupled to the other two mechanical resonators with a rate $J_\mathrm{M}$. An emitter is placed in the central optical cavity and interacts with its optical mode with a coupling rate $g$.}\label{figSuppInfo_Model_3_Cavities}
\end{figure}

The model for the three-cavity case is schematically depicted in fig. \ref{figSuppInfo_Model_3_Cavities}. The three identical optical cavities, denoted left (L), target (T) and right (R) are dispersively coupled at a rate $g_0$ with a separate mechanical resonator, in similar fasion as in the previous section. The three mechanical resonators are assumed identical and with frequency $\Omega_\mathrm{M}$. We describe the three optical cavities with the annihilation operators $\hat{a}_\mathrm{L}$, $\hat{a}_\mathrm{T}$ and $\hat{a}_\mathrm{R}$, and the three resonators with annihilation operators $\hat{b}_\mathrm{L}$, $\hat{b}_\mathrm{T}$ and $\hat{b}_\mathrm{R}$. The optical cavity T is coupled to the cavities L and R with a rate $J$, while the mechanical resonator $\hat{b}_\mathrm{T}$ is coupled with a rate $J_\mathrm{M}$ to the resonators $\hat{b}_\mathrm{R}$ and $\hat{b}_\mathrm{L}$. Finally, a two-level emitter is placed in the target cavity, and interacts with the field $\hat{a}_\mathrm{T}$ with a coupling rate $g$.

The full Hamiltonian reads
\begin{equation}
\label{eq_Hamiltonian3Cavities_1}
\begin{split}
\hat{H} =& \left[\omega_\mathrm{c} - g_0 \left(\hat{b}_\mathrm{L}^\dag +\hat{b}_\mathrm{L} \right) \right] \hat{a}_\mathrm{L}^\dag \hat{a}_\mathrm{L}+ \left[\omega_\mathrm{c} - g_0 \left(\hat{b}_\mathrm{T}^\dag +\hat{b}_\mathrm{T} \right) \right] \hat{a}_\mathrm{T}^\dag \hat{a}_\mathrm{T} + \left[\omega_\mathrm{c} - g_0 \left(\hat{b}_\mathrm{R}^\dag +\hat{b}_\mathrm{R} \right) \right] \hat{a}_\mathrm{R}^\dag \hat{a}_\mathrm{R} + \\ &+\Omega_\mathrm{M} \left(\hat{b}_\mathrm{L}^\dag \hat{b}_\mathrm{L} + \hat{b}_\mathrm{T}^\dag \hat{b}_\mathrm{T} + \hat{b}_\mathrm{R}^\dag \hat{b}_\mathrm{R}\right) + \frac{\omega_\mathrm{A}}{2}\hat{\sigma}_z +\\
&+J \left[\hat{a}_\mathrm{T}^\dag\left(\hat{a}_\mathrm{R}+ \hat{a}_\mathrm{L}\right) + h.c.\right] + J_\mathrm{M} \left[\hat{b}_\mathrm{T}^\dag\left(\hat{b}_\mathrm{R} +  \hat{b}_\mathrm{L} \right) + h.c.\right]+ g\left(\hat{\sigma}_+ \hat{a}_\mathrm{T} + h.c. \right).
\end{split}
\end{equation}
The mechanical part of this Hamiltonian can be diagonalized by introducing the modes
\begin{align}
\label{eq_Hamiltonian3Cavities_CoupledMechModes}
\begin{split}
&\hat{b}_\mathrm{0} = \frac{1}{\sqrt{2}} \left(\hat{b}_\mathrm{L} - \hat{b}_\mathrm{R}\right),\\
&\hat{b}_+ = \frac{1}{\sqrt{2}} \hat{b}_\mathrm{T} + \frac{1}{2}\left(\hat{b}_\mathrm{R} + \hat{b}_\mathrm{L}\right),\\
&\hat{b}_- = -\frac{1}{\sqrt{2}} \hat{b}_\mathrm{T} + \frac{1}{2}\left(\hat{b}_\mathrm{R} + \hat{b}_\mathrm{L}\right).
\end{split}
\end{align}
The transformed Hamiltonian reads
\begin{align}
\label{eq_Hamiltonian3Cavities_2}
\begin{split}
\hat{H} =& \omega_\mathrm{c} \left( \hat{a}_\mathrm{L}^\dag \hat{a}_\mathrm{L}+ \hat{a}_\mathrm{T}^\dag \hat{a}_\mathrm{T} +  \hat{a}_\mathrm{R}^\dag \hat{a}_\mathrm{R}\right)+\Omega_\mathrm{M} \hat{b}_\mathrm{0}^\dag \hat{b}_\mathrm{0}+\left(\Omega_\mathrm{M}+\sqrt{2}J_\mathrm{M}\right) \hat{b}_+^\dag \hat{b}_+ +\left(\Omega_\mathrm{M}-\sqrt{2}J_\mathrm{M} \right) \hat{b}_-^\dag \hat{b}_- + \frac{\omega_\mathrm{A}}{2}\hat{\sigma}_z + \\
&+\frac{g_0}{\sqrt{2}} \hat{x}_0 \left(\hat{a}_\mathrm{R}^\dag \hat{a}_\mathrm{R}-  \hat{a}_\mathrm{L}^\dag \hat{a}_\mathrm{L}\right)+ \frac{g_0}{2} \left(\hat{x}_+ - \hat{x}_-\right) \left(\hat{a}_\mathrm{L}^\dag \hat{a}_\mathrm{L}+  \hat{a}_\mathrm{R}^\dag\hat{a}_\mathrm{R}\right)-\frac{g_0}{2} \left(\hat{x}_+ - \hat{x}_-\right) \hat{a}_\mathrm{T}^\dag \hat{a}_\mathrm{T}+\\
&+J \left[\hat{a}_\mathrm{T}^\dag\left(\hat{a}_\mathrm{R}+ \hat{a}_\mathrm{L}\right) + h.c.\right] + g\left(\hat{\sigma}_+ \hat{a}_\mathrm{T} + h.c. \right),
\end{split}
\end{align}
where we have defined the dimensionless position operators of the mechanical supermodes $\hat{x}_{\pm} = \hat{b}_{\pm}^\dag +\hat{b}_{\pm}$ and $\hat{x}_{0} = \hat{b}_{0}^\dag +\hat{b}_{0}$. We now focus only on the mechanical mode $\hat{b}_\mathrm{0}$ with the assumption that, under resonant conditions, the terms involving the other mechanical modes are negligible (this assumption is numerically verified in section \ref{sec_OtherMechanialModes}). For simplicity, we redefine $\hat{b}_\mathrm{0} \rightarrow \hat{b}$. Moreover, we perform a unitary transformation $\hat{H} \rightarrow  \hat{U}(t) \hat{H} \hat{U}^\dag(t) - i \hat{U}(t) \dfrac{\partial \hat{U}(t)^\dag}{\partial t}$, with $\hat{U}(t) = \exp\left[-i\omega_\mathrm{c} t \left(\hat{a}_\mathrm{L}^\dag \hat{a}_\mathrm{L}+ \hat{a}_\mathrm{T}^\dag \hat{a}_\mathrm{T}+\hat{a}_\mathrm{R}^\dag \hat{a}_\mathrm{R} + \hat{\sigma}_+ \hat{\sigma}_- \right) \right]$. The new Hamiltonian reads
\begin{align}
\label{eq_Hamiltonian3Cavities_3}
\begin{split}
\hat{H} =& -\hat{\Delta}\hat{a}_\mathrm{L}^\dag \hat{a}_\mathrm{L}   +\hat{\Delta}  \hat{a}_\mathrm{R}^\dag \hat{a}_\mathrm{R}+\Omega_\mathrm{M} \hat{b}^\dag \hat{b}+ + \frac{\omega_\mathrm{A}-\omega_\mathrm{c}}{2}\hat{\sigma}_z + \\
&+J \left[\hat{a}_\mathrm{T}^\dag\left(\hat{a}_\mathrm{R}+ \hat{a}_\mathrm{L}\right) + h.c.\right] + g\left(\hat{\sigma}_+ \hat{a}_\mathrm{T} + h.c. \right),
\end{split}
\end{align}
where we defined $\hat{\Delta} = g_0\hat{x}_0/\sqrt{2} $. We note that the selected mechanical mode has equal and opposite dispersive coupling with the optical cavities L and R, at a rate $\pm g_0/\sqrt{2}$, while it does not affect the optical cavity T. We again assume a quasi-static approximation for $\hat{\Delta}$, valid in the limit $J\gg\Omega_\mathrm{M}$, to diagonalize the optical part of the Hamiltonian \citep{SI_ludwig2012}. The three optical supermodes are defined in this case by
\begin{align}
\label{eq_Hamiltonian3Cavities_DefCoupledModes}
\begin{split}
\hat{a}_\mathrm{0} &= -\epsilon \hat{a}_\mathrm{L} + \beta \hat{a}_\mathrm{T} + \epsilon \hat{a}_\mathrm{R}, \\
\hat{a}_\mathrm{+} &= \eta \hat{a}_\mathrm{L} -\epsilon \hat{a}_\mathrm{T} + \mu \hat{a}_\mathrm{R}, \\
\hat{a}_\mathrm{-} &= \mu \hat{a}_\mathrm{L} + \epsilon \hat{a}_\mathrm{T} + \eta \hat{a}_\mathrm{R}, \\
\end{split}
\end{align}
where $\epsilon$, $\beta$, $\mu$ and $\eta$ are operators. In particular,
\begin{equation}
\epsilon = \frac{1}{\sqrt{2+\left(\hat{\Delta}/J\right)^2}}, \;\;\; \beta = \frac{\hat{\Delta}/J}{\sqrt{2+\left(\hat{\Delta}/J\right)^2}}.
\end{equation}
The functions $\eta$ and $\mu$ satisfy $\eta(\hat{\Delta},J) = \mu(-\hat{\Delta},J)$. Their expressions are more complicated and not reported here, as they are not needed in the following. The Hamiltonian in the supermode basis reads
\begin{align}
\label{eq_Hamiltonian3Cavities_4}
\begin{split}
\hat{H} =& \sqrt{2J^2+\hat{\Delta}^2}\left(\hat{a}_\mathrm{+}^\dag \hat{a}_\mathrm{+}   -\hat{a}_\mathrm{-}^\dag \hat{a}_\mathrm{-}\right) +\Omega_\mathrm{M} \hat{b}^\dag \hat{b}+ + \frac{\omega_\mathrm{A}-\omega_\mathrm{c}}{2}\hat{\sigma}_z + \\
&+ g \beta \left(\hat{\sigma}_+ \hat{a}_\mathrm{0} + h.c. \right) + g \epsilon \left[\hat{\sigma}_+ \left(\hat{a}_\mathrm{+} - \hat{a}_\mathrm{-} \right) + h.c. \right].
\end{split}
\end{align}
Up to the first order in $\hat{\Delta}/J$, $\beta$ and $\epsilon$ read
\begin{equation}
\beta = \frac{1}{\sqrt{2}}\frac{\hat{\Delta}}{J} +  \mathcal{O}\left(\frac{\hat{\Delta}}{J}\right)^3, \;\;
\epsilon = \frac{1}{\sqrt{2}} + \mathcal{O}\left(\frac{\hat{\Delta}}{J}\right)^2 .
\end{equation}
By inserting these expansions in the Hamiltonian in eq.~\ref{eq_Hamiltonian3Cavities_4}, and expressing again $\hat{\Delta}$ as a function of $\hat{b}$ and $\hat{b}^\dag$, we get
\begin{align}
\label{eq_Hamiltonian3Cavities_5}
\begin{split}
\hat{H} =& \sqrt{2J^2+\hat{\Delta}^2}\left(\hat{a}_\mathrm{+}^\dag \hat{a}_\mathrm{+}   -\hat{a}_\mathrm{-}^\dag \hat{a}_\mathrm{-}\right) +\Omega_\mathrm{M} \hat{b}^\dag \hat{b}+ + \frac{\omega_\mathrm{A}-\omega_\mathrm{c}}{2}\hat{\sigma}_z + \\
&+ \frac{g g_0}{2J} \left( \hat{b}^\dag +\hat{b} \right) \left(\hat{\sigma}_+ \hat{a}_\mathrm{0} + h.c. \right) + \frac{g}{\sqrt{2}} \left[\hat{\sigma}_+ \left(\hat{a}_\mathrm{+} - \hat{a}_\mathrm{-} \right) + h.c. \right].
\end{split}
\end{align}
The first term of the second row describes the mode field coupling between the emitter, the phonon and the mode $\hat{a}_\mathrm{0}$, with a coupling rate $\gamma=\dfrac{g g_0}{2 J}$. The second term of the second row describes a Rabi interaction between the emitter and the \textit{other} two optical supermodes. As confirmed by the numerical simulations based on the full Hamiltonian in eq.~\ref{eq_Hamiltonian3Cavities_3} (see main text), the effect of these terms is negligible when the spectral separation between the supermodes is much larger than the mechanical frequency ($\sqrt{2} J \gg \Omega_\mathrm{M}$). However, for finite optical linewidths ($\kappa$), the presence of the modes $\hat{a}_\pm$ will introduce additional decay channels for the QE, as explained below.

\section{Emitter decay rate in the three-cavity system}
\begin{figure}[th!] 	
  	\center
    \includegraphics[scale=0.8]{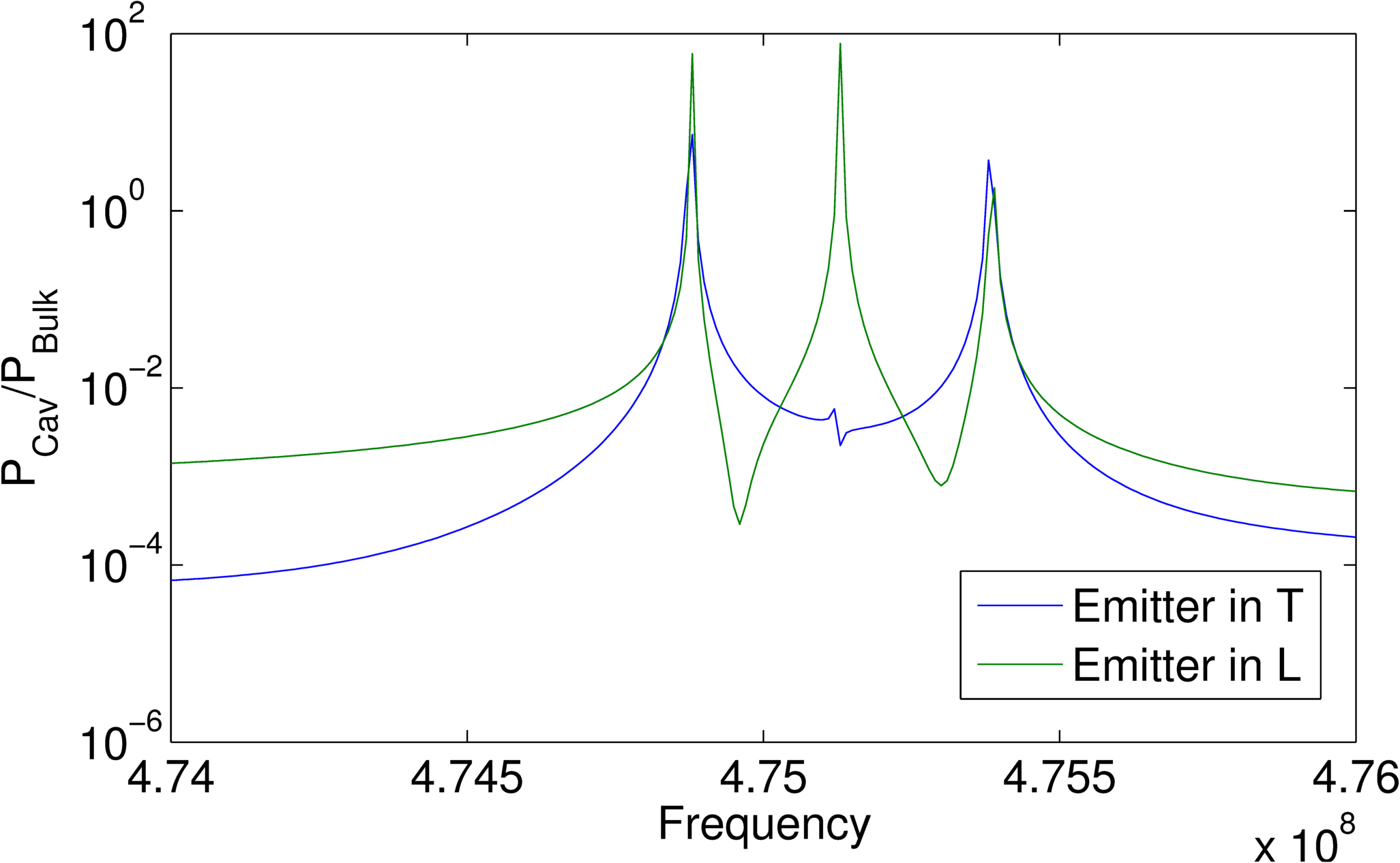}
    \caption{Calculated power emitted by a dipole placed in the target (blue line) and left (green line) cavity of the three-cavity diamond structure discussed in the main text and in sec. \ref{sec_Design} of this SI. The powers ($P_\mathrm{Cav}$) are normalized by the power emitted in bulk diamond ($P_\mathrm{Bulk}$).} \label{figSuppInfo-DecayRateIn3Cav}
\end{figure}

As mentioned in the main text, for the calculation in fig. 3b we assume a conservative value of $\Gamma/2\pi$ = 0.05 GHz for the rate of the emitter decay  into the leaky modes (\textit{i.e.}, all the optical modes different from the three supermodes of the system). This value corresponds to the one measured for an NV center placed in a single defect cavity (similar to the one used here to realize a three-cavity system) and out of resonance with the optical mode \cite{SI_lee2014}. However, significantly smaller radiative decay rates could in principle occur in the proposed structure, as we explain in the following. To evaluate the expected decay rate, we perform finite element method calculations (COMSOL) of the power emitted by an electric dipole (fig. \ref{figSuppInfo-DecayRateIn3Cav}) placed in either the target (blue line) or the left cavity (green line) of the structure considered in the main text. We then normalize the calculated powers by the power emitted by the same dipole in bulk diamond. In this way we obtain a prediction for the radiative decay rate of a QE (with the same position and polarization of the dipole) normalized to the radiative decay rate of the same QE in bulk. In agreement with the spatial patterns of the optical supermodes (see also sec. \ref{sec_Design} of this S.I.), a dipole placed in the lateral cavity can feed all the three optical supermodes, while one that is located in the target cavity does not emit into the mode $\hat{a}_\mathrm{0}$. In particular, the calculation shows that the radiative decay rate of a QE placed in the cavity T is reduced by almost a factor $10^3$ with respect to the decay rate in bulk. As the measured lifetime of an NV center in bulk is about 12 ns \cite{SI_faraon2012coupling}, a radiative decay rate of the order of $\Gamma/2\pi$ = 0.1 MHz is in principle expected for the same emitter placed in the target cavity. We also note that the emission of a dipole in the target cavity is dominated by the coupling with the nonresonant coupled modes for the chosen set of parameters (see sec. \ref{sec_AdditionalDecays}), while the emission into leaky modes is expected to be even smaller. We therefore conclude that the assumed value of 50 MHz is indeed a conservative, upper estimate of the QE decay rate.

\section{Additional radiative losses introduced by the supermodes $\hat{a}_\mathrm{+}$ and $\hat{a}_\mathrm{-}$}
\label{sec_AdditionalDecays}
As shown in the main text, for a proper parameter choice the Hamiltonian of the three-cavity system provides a tripartite interaction between the supermode $\hat{a}_\mathrm{0}$, the QE and the mechanical mode. The interaction between the QE and the modes $\hat{a}_\pm$ is suppressed for large supermode separation ($\sqrt{2} J \gg \Omega_\mathrm{M}$). However, these optical modes are still coupled with the emitter (since their fields at the emitter position is not zero), as shown in fig. \ref {figSuppInfo-DecayRateIn3Cav}. For large optical losses $\kappa > g$, the supermodes $\hat{a}_\pm$ are weakly coupled to the QE and therefore they introduce additional decay channels (denoted $\Gamma^{(\pm)}$). These additional decay channels become relevant when the other radiative decay channel ($\Gamma$) is small and the optical linewidth is significant. This effect is important in understanding some of the features shown in the main text about the swapping fidelity and the ground-state cooling (see fig. 3(c-d) and accompanying discussion in the main text). $\Gamma^{(\pm)}$ can be calculated as follows. If an optical mode with losses $\kappa$ is resonant with a QE and coupled with it at a rate $g' < \kappa$, the QE decays into the optical mode at a rate $4 g'^2/\kappa$ . In the presence of a large spectral detuning $\delta \gg \kappa$, the decay rate is modified into $\dfrac{4 g'^2}{\kappa} \cdot \dfrac{\kappa^2/4}{\delta^2 + \kappa^2/4} \approx \dfrac{g'^2 \kappa}{\delta^2}$ \cite{SI_lodahl2015interfacing}. In our case, $\delta = \sqrt{2} J$ and we replace $g' = g/\sqrt{2}$, consistently with the notation in the main text where $g$ is the coupling between the QE and the uncoupled cavity. Therefore,
\begin{equation}
\Gamma^{(\pm)} = \frac{g^2 \kappa }{4 J^2}.
\end{equation}
With respect to the QE-phonon swapping this additional decay channel is negligible when $\Gamma^{(\pm)} \ll \gamma \sqrt{n_\mathrm{cav}} \Rightarrow \kappa \ll 2 J g_0 \sqrt{n_\mathrm{cav}}/g$, which corresponds to the vertical dashed-dotted line in fig. 3c of the main text.

\section{Selective pumping of the optical mode of interest in the three-cavity system}
In the three-cavity system the mode field coupling is due to a particular optical supermode ($\hat{a}_\mathrm{0}$), which, when the system is not perturbed by the mechanical displacement, features zero electric field in the central cavity and equal and opposite field amplitudes in the lateral cavities (see the first of eqs. \ref{eq_Hamiltonian3Cavities_DefCoupledModes} for $\Delta = 0$). In order to enhance the mode field coupling it is therefore necessary to selectively pump this mode, without feeding the other two optical supermodes. This is possible by pumping the two lateral cavities at a frequency $\omega_\mathrm{c}$ with equal field amplitude $\mathcal{E}$ and opposite phases: excitation of modes $\hat{a}_\pm$ is then symmetry-forbidden. Focusing only on the optical part of the Hamiltonian, we have
\begin{align}
\label{eq_SelectivePumping_1}
\begin{split}
\hat{H}_{opt} =J \left[\hat{a}_\mathrm{T}^\dag\left(\hat{a}_\mathrm{R}+ \hat{a}_\mathrm{L}\right) + h.c.\right] + \mathcal{E}\left(\hat{a}^\dag_R - \hat{a}^\dag_L + h.c.\right),
\end{split}
\end{align}
where we switched to a frame rotating at frequency $\omega_\mathrm{c}$. By writing the equations of motion for the three cavity field amplitudes, and assuming equal cavity losses $\kappa$, we can calculate the steady state population under this pumping scheme, namely $\bar{a}_R = -\bar{a}_L = 2 \mathcal{E}/\kappa$ and $\bar{a}_T = 0$. By comparing these results with eqs. \ref{eq_Hamiltonian3Cavities_DefCoupledModes} for $\Delta = 0$, we see that only the mode $\hat{a}_\mathrm{0}$ is fed, while the population of the other two modes remains strictly zero. Interestingly, in the limit of validity of the coupled mode theory, this result (\textit{i.e.} the zero population of the supermodes $\hat{a}_\mathrm{+}$ and $\hat{a}_\mathrm{-}$) is independent of $\kappa$.

\section{Three-cavity system Hamiltonian at large driving}
In the previous section we explained how the mode $\hat{a}_\mathrm{0}$ can be selectively pumped. However, solving a full quantum model (eq. \ref{eq_Hamiltonian3Cavities_3}) with a large pumping  of the optical cavities  is computationally extremely challenging, because of the large Fock space dimensions required. We therefore solve a transformed version of the Hamiltonian in eq. \ref{eq_Hamiltonian3Cavities_3}, obtained through a displacement of the cavity operators. That is, we replace the operators by the sum of a steady-state amplitude and a fluctuating operator, \textit{i.e.} $\hat{a}_i = \bar{a}_i + \delta \hat{a}_i$, where $i=$ L,R,T. The transformed Hamiltonian reads
\begin{align}
\label{eq_Hamiltonian3Cavities_Linearized}
\begin{split}
\hat{H}_{lin} =& -\hat{\Delta} \delta\hat{a}^\dag_L  \delta\hat{a}_\mathrm{L} + \hat{\Delta}  \delta\hat{a}^\dag_R  \delta\hat{a}_\mathrm{R} + \Omega_\mathrm{M} \hat{b}^\dag \hat{b} + \frac{\omega_\mathrm{A}-\omega_\mathrm{c}}{2}\hat{\sigma}_z + J \left( \delta\hat{a}^\dag_T \delta\hat{a}_\mathrm{L} + h.c. +  \delta\hat{a}^\dag_T  \delta\hat{a}_\mathrm{R} + h.c. \right) + \\ &+ g\left(  \delta\hat{a}_\mathrm{T} \hat{\sigma}_+ + h.c. \right) + \frac{g_0}{\sqrt{2}} \sqrt{\frac{n_\mathrm{cav}}{2}} \left(\hat{b}^\dag + \hat{b} \right) \left[ \left(  \delta\hat{a}^\dag_L +  \delta\hat{a}_\mathrm{L} \right) + \left( \delta\hat{a}^\dag_R +\delta\hat{a}_\mathrm{R} \right)\right ]
\end{split}
\end{align}
where we defined $\sqrt{n_\mathrm{cav}}=\sqrt{2}|\bar{a}_R| = \sqrt{2}|\bar{a}_L|$ and we have used the fact that, with the pumping scheme described in the previous section, $\bar{a}_R = -\bar{a}_L$ and $\bar{a}_T$ = 0. We note that this Hamiltonian is still equivalent to the original one, since no terms have been neglected. The Hamiltonian has been solved numerically (with the method discussed in sec. \ref{sec_MasterEQ}) by either setting $n_\mathrm{cav}$ = 0 (for the case in which the mode $\hat{a}_\mathrm{0}$ is not externally pumped) or setting a fixed value of $n_\mathrm{cav}$ (continuous optical pumping of the system). To address the case of a square-pulse excitation, we first calculate the time evolution of the (classical) cavity fields amplitudes upon the external pumping described above,  neglecting the presence of the QE and the mechanical mode. From the amplitudes $\bar{a}_R(t)$ and $\bar{a}_L(t)$, we calculate the population of the mode $\hat{a}_\mathrm{0}$ ($n_\mathrm{cav}(t)$), which is used as a time-dependent parameter in solving the Hamiltonian in eq. \ref{eq_Hamiltonian3Cavities_Linearized}.

\section{Influence of cavity population and dephasing on the QE-Phonon swapping fidelity}
\begin{figure}[th!] 	
  	\center
    \includegraphics[scale=0.75]{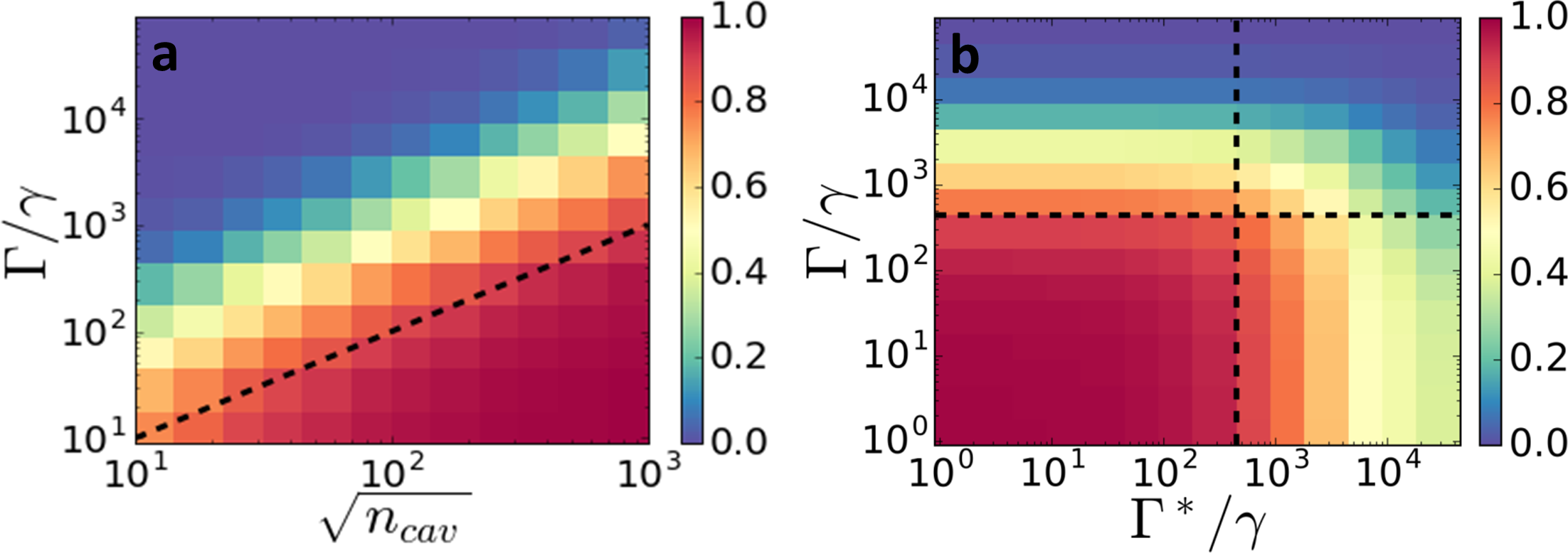}
    \caption{(a) Fidelity of the emitter-phonon swapping versus the emitter decay rate $\Gamma$ and the population of the mode $\hat{a}_\mathrm{0}$. The dashed black line indicates the condition $\Gamma/\gamma=\sqrt{n_\mathrm{cav}}$. (b) Fidelity of the emitter-phonon swapping versus the emitter decay rate $\Gamma$ and the emitter pure dephasing rate $\Gamma^*$. The horizontal and vertical dashed black line indicate the conditions $\Gamma/\gamma =\sqrt{n_\mathrm{cav}}$ and $\Gamma^*/\gamma =\sqrt{n_\mathrm{cav}}$, respectively. }\label{figSuppInfo-FidelityVSncav_dephasing}
\end{figure}
As mentioned in the main text, the mode field coupling rate can be enhanced by increasing the population of the mode $\hat{a}_\mathrm{0}$. This is beneficial for, $\textit{e.g.}$, the emitter-phonon swapping fidelity. The colormap in fig. \ref{figSuppInfo-FidelityVSncav_dephasing}a shows the swapping fidelity (calculated with the linearized Hamiltonian of the three-cavity system, eq. \ref{eq_Hamiltonian3Cavities_Linearized}) versus the emitter decay rate $\Gamma$ and the population of the mode $\hat{a}_\mathrm{0}$. All the other parameters are the same as in the main text. As expected, near-unity fidelity is obtained  when $\gamma\sqrt{n_\mathrm{cav}}>\Gamma$.

Due to the coherent nature of the MFC, pure dephasing of the QE is expected to further decrease the swapping fidelity.  As shown in fig. \ref{figSuppInfo-FidelityVSncav_dephasing}b, the QE decay $\Gamma$ and pure dephasing rates $\Gamma^*$ play a similar role in determining the swapping fidelity, and $\Gamma^* \ll \gamma \sqrt{n_{\mathrm{cav}}}$ is required as well to achieve large fidelity. We note that for the system considered for the calculations shown in this work (NV centers in diamond), the pure dephasing rate is typically of the same order of magnitude of the decay rate in bulk \cite{SI_robledo2010control}.

\section{Role of the other two mechanical supermodes}
\label{sec_OtherMechanialModes}
\begin{figure}[th!] 	
  	\center
    \includegraphics[scale=0.9]{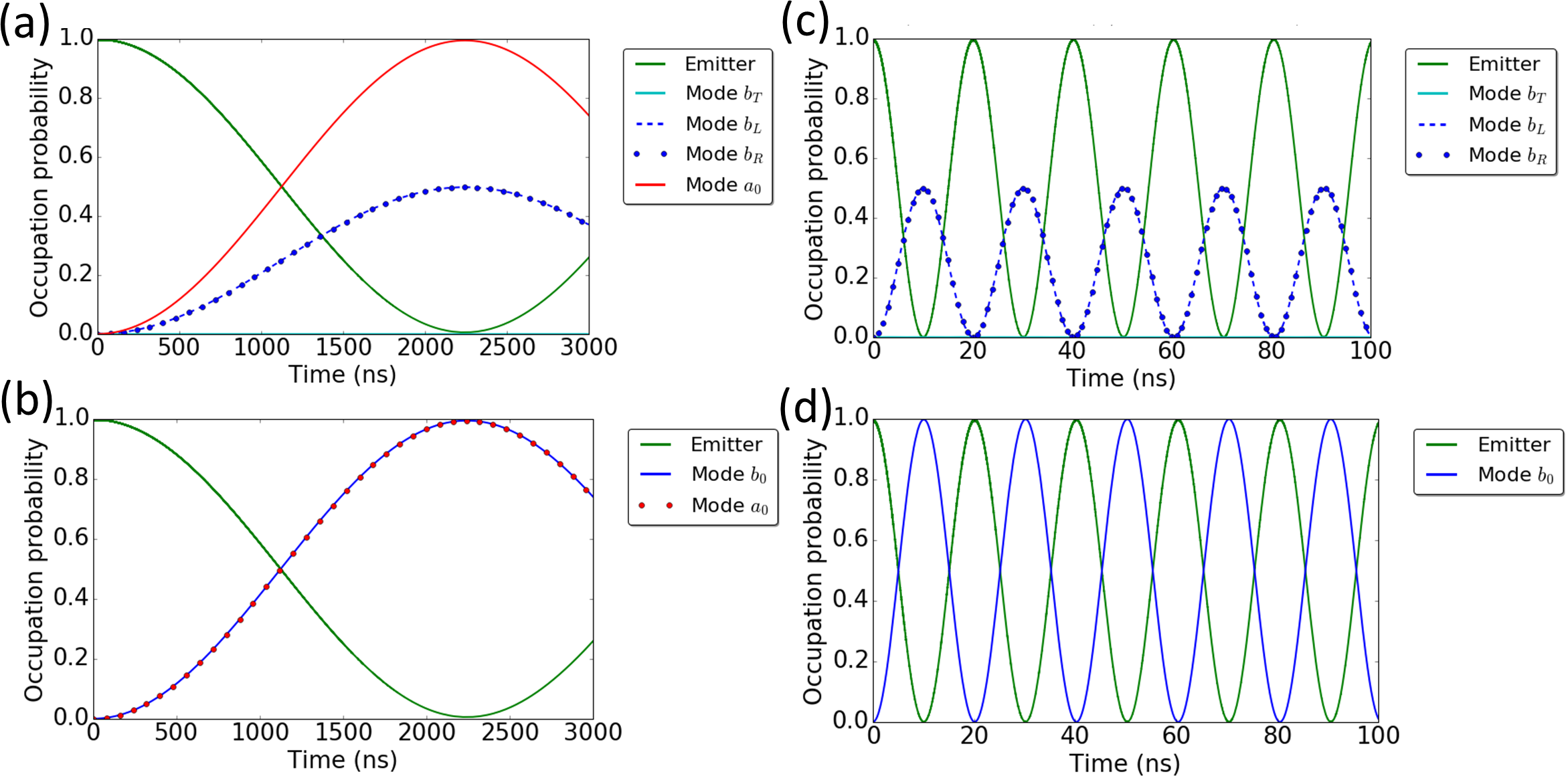}
    \caption{Numerical calculation of the full Hamiltonian in eq.  \ref{eq_Hamiltonian3Cavities_1}, \textit{i.e.} considering all mechanical modes. The parameters of the system are the same as in fig. 3a of the main text. (a-b) Vacuum oscillations of the system (\textit{i.e.} the optical cavities are not pumped). (a) The mechanical modes $\hat{b}_\mathrm{L}$ and $\hat{b}_\mathrm{R}$ are equally populated (blue solid and blue dotted lines) and reach a maximum occupation probability of 0.5, while the mechanical mode $\hat{b}_\mathrm{T}$ is not populated (cyan solid line). (b) When switching to the supermode basis (through eqs. \ref{eq_Hamiltonian3Cavities_CoupledMechModes}), the mode $\hat{b}_\mathrm{0}$ behaves exactly as shown in previous calculations in which the other supermodes were neglected.  (c-d) Same as in panels (a-b) but with an external pumping of the optical cavities, such that the optical supermode $\hat{a}_\mathrm{0}$ contains $n=5\cdot 10^4$ photons.} \label{figSuppInfo_Other2MechanicalModes}
\end{figure}
\begin{figure}[th!] 	
  	\center
    \includegraphics[width=\linewidth]{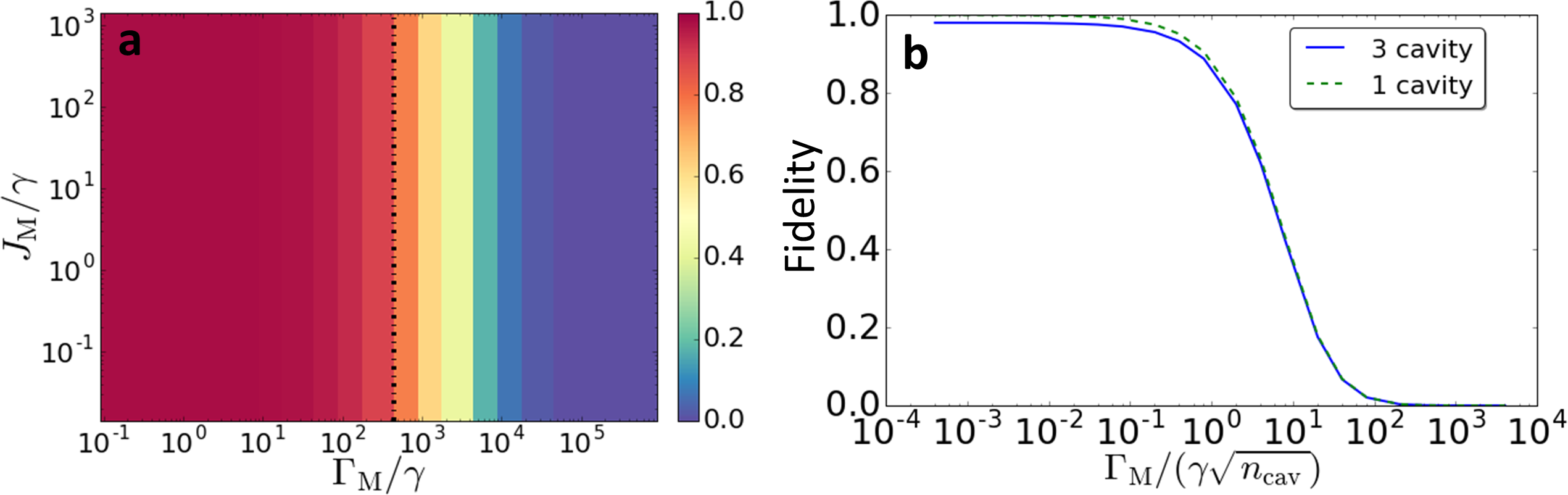}
    \caption{(a) QE-phonon swapping fidelity versus the mechanical interaction $J_\mathrm{M}$ and the mechanical losses $\Gamma_\mathrm{M}$. All the other parameters are the same as in the main text and the QE losses have been neglected. The system is pumped with $n_\mathrm{cav}=5\cdot10^4$. The dashed-dotted vertical line indicates the condition $\Gamma_\mathrm{M}=\gamma \sqrt{n_\mathrm{cav}}$. (b) Horizontal cut of panel (a) for the lowest value of $J_\mathrm{M}$ (solid blue line) compared with the fidelity calculated for the same mechanical losses but in the MFC model with one cavity and one resonator (green dashed line), see eq. 2 of main text.} \label{figSuppInfo-FidelityVS_JM_GammaM}
\end{figure}

In the main text, and in the previous sections of this supplemental information, we assumed that, once the mechanical modes are hybridized, only the mechanical supermode of interest is relevant for the system dynamics. To verify this assumption, we perform additional numerical calculations based on the Hamiltonian in eq. \ref{eq_Hamiltonian3Cavities_1}, \textit{i.e.} by considering explicitly the three mechanical uncoupled modes. We use the same parameters as in the main text, and $J_\mathrm{M}/(2\pi) = 50$ MHz. Initially, we assume no mechanical losses ($\Gamma_\mathrm{M}=0$). After calculating the time evolution of the modes $\hat{b}_\mathrm{L}$, $\hat{b}_\mathrm{R}$ and $\hat{b}_\mathrm{T}$, we derive the time evolution of the three mechanical supermodes based on eqs. \ref{eq_Hamiltonian3Cavities_CoupledMechModes}.  As shown in fig. \ref{figSuppInfo_Other2MechanicalModes}, both in the unpumped and the pumped case the dynamics of the mechanical mode $\hat{b}_\mathrm{0}$ matches that found previously, where the other mechanical modes were neglected. The population of the other two mechanical supermodes ($\hat{b}_+$ and $\hat{b}_-$) is always zero (not shown in these plots). This is due to the fact that the other two mechanical supermodes interact with the optical cavities in a way which does not lead to any tripartite interaction (see eq. \ref{eq_Hamiltonian3Cavities_2}), and their presence is therefore negligible for proper choice of frequencies.
Indeed, we also verified (fig. \ref{figSuppInfo-FidelityVS_JM_GammaM}a) that the swapping fidelity does not depend on the mechanical interaction $J_\mathrm{M}$ even for large mechanical losses $\Gamma_\mathrm{M} \gg J_\mathrm{M}$. Decrease of the fidelity is only observed if the mechanical losses become larger than $\gamma \sqrt{n_\mathrm{cav}}$. The fact that the the presence of the other two mechanical supermodes can be completely neglected is also confirmed by fig. \ref{figSuppInfo-FidelityVS_JM_GammaM}b: here, we compare the swapping fidelity versus $\Gamma_\mathrm{M}$ for the case of a three-cavity system (\textit{i.e.} a horizontal cut of panel a) with the same graph calculated for a system composed by one optical cavity and one mechanical resonator, where the MFC is introduced \textit{ad-hoc} (eq. 2 of the main text). The identical behaviours indicate that a large $\Gamma_\mathrm{M}$ introduces only additional losses but does not make the system interact with the other mechanical supermodes. The small deviation in the fidelity for $\Gamma_\mathrm{M}/(\gamma \sqrt{n_\mathrm{cav}})<1$ is due to the additional losses present in the three-cavity system due to the optical modes $\hat{a}_\pm$.

\newpage
\section{Design of the proposed structure}
\label{sec_Design}
\begin{figure}[th!] 	
  	\center
    \includegraphics[scale=0.63]{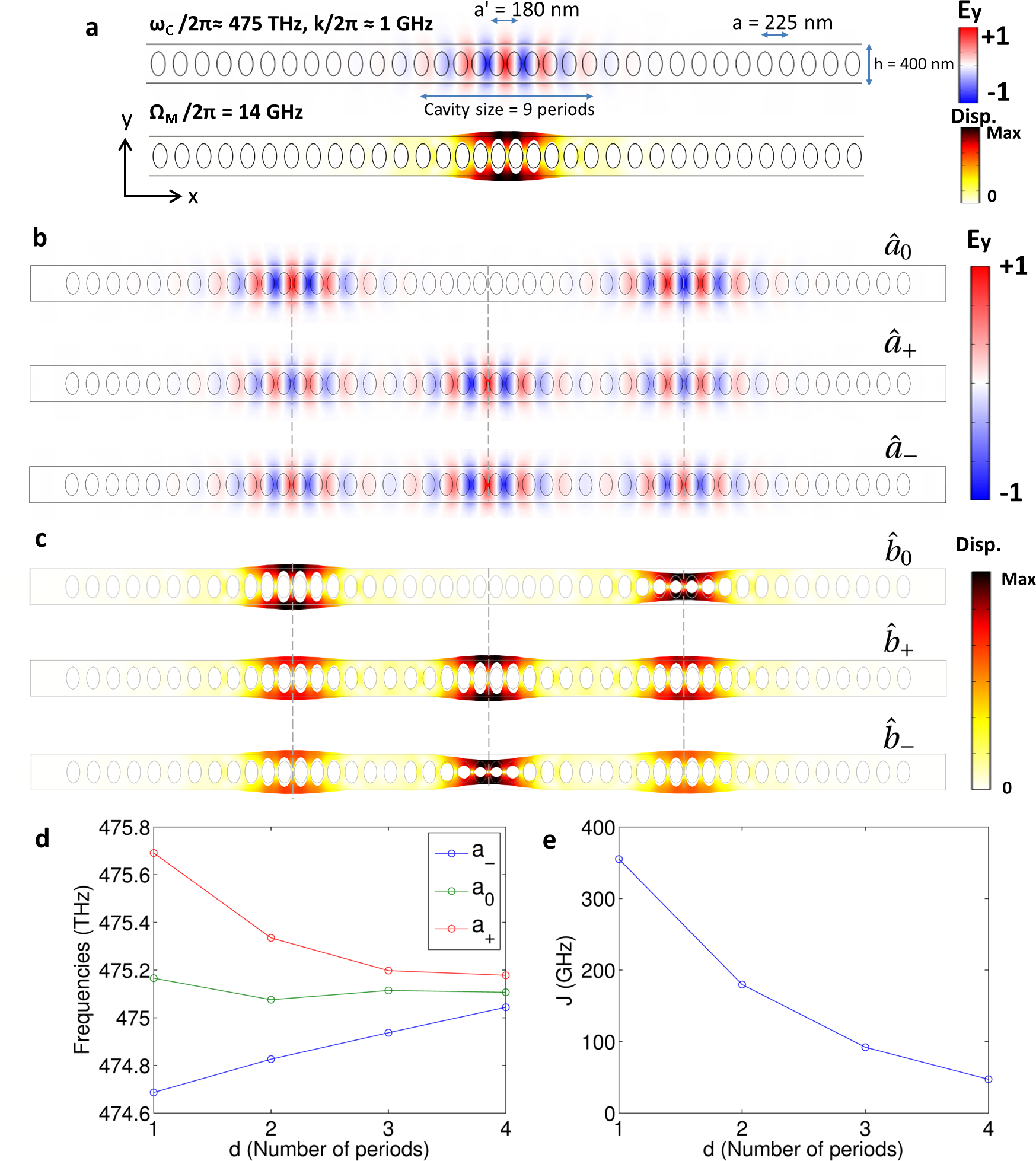}
    \caption{Numerical investigation of the proposed structure (see text for details). (a) Electric field pattern (top) and mechanical displacement pattern (bottom) of the colocalized optical and mechanical mode for a single cavity. The $y$-component of the electric field is shown. The mode frequencies and the optical losses are indicated in the figure. (b-c) Three identical cavities on a nanobeam, separated by 2 periods. The position of the three cavities is marked by the vertical grey dashed lines. (b) Electric field pattern of the three optical supermodes when the mechanical mode is at rest. (c) Displacement pattern of the three mechanical supermodes. (d) Frequencies of the three optical supermodes as a function of the cavities separation (\textit{i.e.} number of periods). (e) Values of the optical interaction $J$ (extrapolated from the plot in panel (d), see text) versus the cavity separation.}\label{figSuppInfo_DesignCavities_CoupledModes}
\end{figure}
The structure proposed in the main text is composed of three in-line  defect cavities in a photonic crystal diamond nanobeam. Simulations have been performed with a finite element method commercial software (COMSOL). We first designed the single cavity structure (fig. \ref{figSuppInfo_DesignCavities_CoupledModes}a), which is realized by quadratically tapering the lattice constant across 9 lattice periods, from the unperturbed value of $a = 225$~nm to the central value of $a'=0.8a = 180$~nm. The design is inspired by the one proposed recently by Lee \textit{et al.}\cite{SI_lee2014}. The holes have elliptical shape, with semiaxes equal to $a_x = 0.32a$ and $a_y = 0.55a$. The width of the nanobeam (along the $y$-direction) is $w = 400$~nm, and the thickness is 200~nm. This structure supports an optical mode with frequency $\omega_\mathrm{c}/2\pi= 475$~THz and $Q \approx 3 \cdot 10^5$ (fig. \ref{figSuppInfo_DesignCavities_CoupledModes}a, top panel), and a co-localized mechanical mode with frequency $\Omega_\mathrm{M}/2\pi = 14$~GHz (fig. \ref{figSuppInfo_DesignCavities_CoupledModes}a, bottom panel). The dispersive optomechanical coupling between the optical and mechanical modes has been calculated by evaluating the moving-boundary and photoelastic contribution separately (as discussed in ref. \citenum{SI_burek2015diamond}) and amounts to  $g_0/2\pi \approx 4$~MHz. We notice that we did not perform any systematic optimization on our design, and therefore the optical Q factor and optomechanical coupling rate could be further increased by carefully optimizing the design parameters (\textit{e.g.}, by tapering also the holes' semiaxes) as discussed by other authors \cite{SI_burek2015diamond}.

After having chosen a design for the single cavity, we considered three identical defect cavities on the same nanobeam, separated by an equal distance $d$, defined as the number of unperturbed periods between two adjacent cavities. For each distance $d$, we calculated the optical and mechanical supermodes. Figure~\ref{figSuppInfo_DesignCavities_CoupledModes}b shows the $E_y$ field component for the three optical supermodes. Note that the optical mode of interest, $\hat{a}_\mathrm{0}$, has no field in the central cavity, differently from the other two modes. The modes $\hat{a}_\mathrm{+}$ and $\hat{a}_\mathrm{-}$ differ in the relative sign between the field in the lateral cavities and the field in the central cavity, as expected from the results of the coupled-mode theory (eq. \ref{eq_Hamiltonian3Cavities_DefCoupledModes}). Figure~\ref{figSuppInfo_DesignCavities_CoupledModes}c shows the displacement pattern of three mechanical supermodes. Similarly to the optical case, the mechanical supermodes $\hat{b}_+$ and $\hat{b}_-$ have opposing oscillation phase in the central and the lateral cavities. Figure~\ref{figSuppInfo_DesignCavities_CoupledModes}d shows the frequencies of the three optical supermodes versus the cavity separation $d$. The frequency of the mode $\hat{a}_\mathrm{0}$ is expected to be independent of the cavity interaction $J$ (and therefore the cavity distances). The small frequency deviations observed can be due to either the finite mesh size (which makes the cavities slightly different from each other) or to the breakdown of the coupled-mode theory for very short cavity distances. We notice that this does not affect our theoretical model, since the mechanical movement does not change the rate $J$. In fig. \ref{figSuppInfo_DesignCavities_CoupledModes}e we show the estimated optical interaction rate $J$ as a function of the inter-cavity distance. For each distance $d$, $J$ has been estimated by  the formula $|\omega_+ - \omega_-| = 2 \sqrt{2} J$.

In fig. 2I of the main text we show the expected pattern of the mode $\hat{a}_\mathrm{0}$ for the case in which the system is mechanically perturbed, such that the lateral cavities are detuned by $\Delta/J=0.5$. The shown electric field pattern has been calculated analytically from the formula 
\begin{equation}
E_{0,\Delta}(x,y) = C_0(\Delta) E_0(x,y) + C_+(\Delta) E_+(x,y) + C_-(\Delta) E_-(x,y), 
\end{equation}
where $E_i(x,y)$, with $i=\{0,+,-\}$, are the electric field patterns of the supermodes when the mechanical mode is at rest (fig. \ref{figSuppInfo_DesignCavities_CoupledModes}b), and the coefficients $C_i(\Delta)$ are calculated analytical from the coupled mode theory.
\section{Tripartite interaction in a Fabry-Perot cavity}
The interaction described in this work relies on a variation of the electric field at the emitter position upon a mechanical displacement. A small field variation occurs in principle also in a simple Fabry-Perot (FP) cavity when one of the mirrors is displaced by a mechanical resonator \citep{SI_barzanjeh2011,SI_chang2009,SI_wang2010}. 
However, the interaction strength in this case is much smaller than the one obtained in the two and three cavity systems. In particular, for an emitter placed in a node of the $n$-th FP mode, the tripartite coupling rate  reads $\gamma_n =  \pi g g_{0}/\omega_1$,
where $g$ is the maximum Rabi coupling between the QE and the $n$-th mode, $g_0$ is the dispersive coupling induced by the mirror movement and $\omega_1$ is the fundamental cavity frequency. The coupling rate bears similarity with the one derived for the two- and three-cavity systems, but with the important difference that the role of the intercavity interaction rate $J$ is taken by $\omega_1$, which, in the FP cavity, corresponds also to the frequency spacing between the unperturbed modes. This, in the visible and near-IR regime, severely limits the achievable values of $\gamma_n$, especially since any effort to reduce $\omega_1$ reduces both $g$ and $g_0$.


\begin{thebibliography}{27}%
\makeatletter
\providecommand \@ifxundefined [1]{%
 \@ifx{#1\undefined}
}%
\providecommand \@ifnum [1]{%
 \ifnum #1\expandafter \@firstoftwo
 \else \expandafter \@secondoftwo
 \fi
}%
\providecommand \@ifx [1]{%
 \ifx #1\expandafter \@firstoftwo
 \else \expandafter \@secondoftwo
 \fi
}%
\providecommand \natexlab [1]{#1}%
\providecommand \enquote  [1]{``#1''}%
\providecommand \bibnamefont  [1]{#1}%
\providecommand \bibfnamefont [1]{#1}%
\providecommand \citenamefont [1]{#1}%
\providecommand \href@noop [0]{\@secondoftwo}%
\providecommand \href [0]{\begingroup \@sanitize@url \@href}%
\providecommand \@href[1]{\@@startlink{#1}\@@href}%
\providecommand \@@href[1]{\endgroup#1\@@endlink}%
\providecommand \@sanitize@url [0]{\catcode `\\12\catcode `\$12\catcode
  `\&12\catcode `\#12\catcode `\^12\catcode `\_12\catcode `\%12\relax}%
\providecommand \@@startlink[1]{}%
\providecommand \@@endlink[0]{}%
\providecommand \url  [0]{\begingroup\@sanitize@url \@url }%
\providecommand \@url [1]{\endgroup\@href {#1}{\urlprefix }}%
\providecommand \urlprefix  [0]{URL }%
\providecommand \Eprint [0]{\href }%
\providecommand \doibase [0]{http://dx.doi.org/}%
\providecommand \selectlanguage [0]{\@gobble}%
\providecommand \bibinfo  [0]{\@secondoftwo}%
\providecommand \bibfield  [0]{\@secondoftwo}%
\providecommand \translation [1]{[#1]}%
\providecommand \BibitemOpen [0]{}%
\providecommand \bibitemStop [0]{}%
\providecommand \bibitemNoStop [0]{.\EOS\space}%
\providecommand \EOS [0]{\spacefactor3000\relax}%
\providecommand \BibitemShut  [1]{\csname bibitem#1\endcsname}%
\let\auto@bib@innerbib\@empty
\bibitem [{\citenamefont {Aspelmeyer}\ \emph {et~al.}(2014)\citenamefont
  {Aspelmeyer}, \citenamefont {Kippenberg},\ and\ \citenamefont
  {Marquardt}}]{aspelmeyer2014cavity}%
  \BibitemOpen
  \bibfield  {author} {\bibinfo {author} {\bibfnamefont {M.}~\bibnamefont
  {Aspelmeyer}}, \bibinfo {author} {\bibfnamefont {T.~J.}\ \bibnamefont
  {Kippenberg}}, \ and\ \bibinfo {author} {\bibfnamefont {F.}~\bibnamefont
  {Marquardt}},\ }\href@noop {} {\bibfield  {journal} {\bibinfo  {journal}
  {Rev. of Mod. Phys.}\ }\textbf {\bibinfo {volume} {86}},\ \bibinfo {pages}
  {1391} (\bibinfo {year} {2014})}\BibitemShut {NoStop}%
\bibitem [{\citenamefont {O'Connell}\ \emph {et~al.}(2010)\citenamefont
  {O'Connell}, \citenamefont {Hofheinz}, \citenamefont {Ansmann}, \citenamefont
  {Bialczak}, \citenamefont {Lenander}, \citenamefont {Lucero}, \citenamefont
  {Neeley}, \citenamefont {Sank}, \citenamefont {Wang}, \citenamefont {Weides},
  \citenamefont {Wenner}, \citenamefont {Martinis},\ and\ \citenamefont
  {Cleland}}]{o2010quantum}%
  \BibitemOpen
  \bibfield  {author} {\bibinfo {author} {\bibfnamefont {A.~D.}\ \bibnamefont
  {O'Connell}}, \bibinfo {author} {\bibfnamefont {M.}~\bibnamefont {Hofheinz}},
  \bibinfo {author} {\bibfnamefont {M.}~\bibnamefont {Ansmann}}, \bibinfo
  {author} {\bibfnamefont {R.~C.}\ \bibnamefont {Bialczak}}, \bibinfo {author}
  {\bibfnamefont {M.}~\bibnamefont {Lenander}}, \bibinfo {author}
  {\bibfnamefont {E.}~\bibnamefont {Lucero}}, \bibinfo {author} {\bibfnamefont
  {M.}~\bibnamefont {Neeley}}, \bibinfo {author} {\bibfnamefont
  {D.}~\bibnamefont {Sank}}, \bibinfo {author} {\bibfnamefont {H.}~\bibnamefont
  {Wang}}, \bibinfo {author} {\bibfnamefont {M.}~\bibnamefont {Weides}},
  \bibinfo {author} {\bibfnamefont {J.}~\bibnamefont {Wenner}}, \bibinfo
  {author} {\bibfnamefont {J.}~\bibnamefont {Martinis}}, \ and\ \bibinfo
  {author} {\bibfnamefont {A.~N.}\ \bibnamefont {Cleland}},\ }\href@noop {}
  {\bibfield  {journal} {\bibinfo  {journal} {Nature}\ }\textbf {\bibinfo
  {volume} {464}},\ \bibinfo {pages} {697} (\bibinfo {year}
  {2010})}\BibitemShut {NoStop}%
\bibitem [{\citenamefont {Lecocq}\ \emph {et~al.}(2015)\citenamefont {Lecocq},
  \citenamefont {Teufel}, \citenamefont {Aumentado},\ and\ \citenamefont
  {Simmonds}}]{lecocq2015resolving}%
  \BibitemOpen
  \bibfield  {author} {\bibinfo {author} {\bibfnamefont {F.}~\bibnamefont
  {Lecocq}}, \bibinfo {author} {\bibfnamefont {J.~D.}\ \bibnamefont {Teufel}},
  \bibinfo {author} {\bibfnamefont {J.}~\bibnamefont {Aumentado}}, \ and\
  \bibinfo {author} {\bibfnamefont {R.~W.}\ \bibnamefont {Simmonds}},\
  }\href@noop {} {\bibfield  {journal} {\bibinfo  {journal} {Nat. Phys.}\
  }\textbf {\bibinfo {volume} {11}},\ \bibinfo {pages} {635} (\bibinfo {year}
  {2015})}\BibitemShut {NoStop}%
\bibitem [{\citenamefont {Pirkkalainen}\ \emph {et~al.}(2015)\citenamefont
  {Pirkkalainen}, \citenamefont {Cho}, \citenamefont {Massel}, \citenamefont
  {Tuorila}, \citenamefont {Heikkil{\"a}}, \citenamefont {Hakonen},\ and\
  \citenamefont {Sillanp{\"a}{\"a}}}]{pirkkalainen2015cavity}%
  \BibitemOpen
  \bibfield  {author} {\bibinfo {author} {\bibfnamefont {J.-M.}\ \bibnamefont
  {Pirkkalainen}}, \bibinfo {author} {\bibfnamefont {S.}~\bibnamefont {Cho}},
  \bibinfo {author} {\bibfnamefont {F.}~\bibnamefont {Massel}}, \bibinfo
  {author} {\bibfnamefont {J.}~\bibnamefont {Tuorila}}, \bibinfo {author}
  {\bibfnamefont {T.}~\bibnamefont {Heikkil{\"a}}}, \bibinfo {author}
  {\bibfnamefont {P.}~\bibnamefont {Hakonen}}, \ and\ \bibinfo {author}
  {\bibfnamefont {M.}~\bibnamefont {Sillanp{\"a}{\"a}}},\ }\href@noop {}
  {\bibfield  {journal} {\bibinfo  {journal} {Nat. Comm.}\ }\textbf {\bibinfo
  {volume} {6}} (\bibinfo {year} {2015})}\BibitemShut {NoStop}%
\bibitem [{\citenamefont {Golter}\ \emph {et~al.}(2016)\citenamefont {Golter},
  \citenamefont {Oo}, \citenamefont {Amezcua}, \citenamefont {Stewart},\ and\
  \citenamefont {Wang}}]{golter2016}%
  \BibitemOpen
  \bibfield  {author} {\bibinfo {author} {\bibfnamefont {D.~A.}\ \bibnamefont
  {Golter}}, \bibinfo {author} {\bibfnamefont {T.}~\bibnamefont {Oo}}, \bibinfo
  {author} {\bibfnamefont {M.}~\bibnamefont {Amezcua}}, \bibinfo {author}
  {\bibfnamefont {K.~A.}\ \bibnamefont {Stewart}}, \ and\ \bibinfo {author}
  {\bibfnamefont {H.}~\bibnamefont {Wang}},\ }\href@noop {} {\bibfield
  {journal} {\bibinfo  {journal} {Phys. Rev. Lett.}\ }\textbf {\bibinfo
  {volume} {116}},\ \bibinfo {pages} {143602} (\bibinfo {year}
  {2016})}\BibitemShut {NoStop}%
\bibitem [{\citenamefont {Ramos}\ \emph {et~al.}(2013)\citenamefont {Ramos},
  \citenamefont {Sudhir}, \citenamefont {Stannigel}, \citenamefont {Zoller},\
  and\ \citenamefont {Kippenberg}}]{ramos2013}%
  \BibitemOpen
  \bibfield  {author} {\bibinfo {author} {\bibfnamefont {T.}~\bibnamefont
  {Ramos}}, \bibinfo {author} {\bibfnamefont {V.}~\bibnamefont {Sudhir}},
  \bibinfo {author} {\bibfnamefont {K.}~\bibnamefont {Stannigel}}, \bibinfo
  {author} {\bibfnamefont {P.}~\bibnamefont {Zoller}}, \ and\ \bibinfo {author}
  {\bibfnamefont {T.~J.}\ \bibnamefont {Kippenberg}},\ }\href@noop {}
  {\bibfield  {journal} {\bibinfo  {journal} {Phys. Rev. Lett.}\ }\textbf
  {\bibinfo {volume} {110}},\ \bibinfo {pages} {193602} (\bibinfo {year}
  {2013})}\BibitemShut {NoStop}%
\bibitem [{\citenamefont {Wilson-Rae}\ \emph {et~al.}(2004)\citenamefont
  {Wilson-Rae}, \citenamefont {Zoller},\ and\ \citenamefont
  {Imamoglu}}]{wilson2004}%
  \BibitemOpen
  \bibfield  {author} {\bibinfo {author} {\bibfnamefont {I.}~\bibnamefont
  {Wilson-Rae}}, \bibinfo {author} {\bibfnamefont {P.}~\bibnamefont {Zoller}},
  \ and\ \bibinfo {author} {\bibfnamefont {A.}~\bibnamefont {Imamoglu}},\
  }\href@noop {} {\bibfield  {journal} {\bibinfo  {journal} {Phys. Rev. Lett.}\
  }\textbf {\bibinfo {volume} {92}},\ \bibinfo {pages} {075507} (\bibinfo
  {year} {2004})}\BibitemShut {NoStop}%
\bibitem [{\citenamefont {Restrepo}\ \emph {et~al.}(2014)\citenamefont
  {Restrepo}, \citenamefont {Ciuti},\ and\ \citenamefont
  {Favero}}]{restrepo2014}%
  \BibitemOpen
  \bibfield  {author} {\bibinfo {author} {\bibfnamefont {J.}~\bibnamefont
  {Restrepo}}, \bibinfo {author} {\bibfnamefont {C.}~\bibnamefont {Ciuti}}, \
  and\ \bibinfo {author} {\bibfnamefont {I.}~\bibnamefont {Favero}},\
  }\href@noop {} {\bibfield  {journal} {\bibinfo  {journal} {Phys. Rev. Lett.}\
  }\textbf {\bibinfo {volume} {112}},\ \bibinfo {pages} {013601} (\bibinfo
  {year} {2014})}\BibitemShut {NoStop}%
\bibitem [{\citenamefont {Barzanjeh}\ \emph {et~al.}(2011)\citenamefont
  {Barzanjeh}, \citenamefont {Naderi},\ and\ \citenamefont
  {Soltanolkotabi}}]{barzanjeh2011}%
  \BibitemOpen
  \bibfield  {author} {\bibinfo {author} {\bibfnamefont {S.}~\bibnamefont
  {Barzanjeh}}, \bibinfo {author} {\bibfnamefont {M.~H.}\ \bibnamefont
  {Naderi}}, \ and\ \bibinfo {author} {\bibfnamefont {M.}~\bibnamefont
  {Soltanolkotabi}},\ }\href@noop {} {\bibfield  {journal} {\bibinfo  {journal}
  {Phys. Rev. A}\ }\textbf {\bibinfo {volume} {84}},\ \bibinfo {pages} {063850}
  (\bibinfo {year} {2011})}\BibitemShut {NoStop}%
\bibitem [{\citenamefont {Chang}\ \emph {et~al.}(2009)\citenamefont {Chang},
  \citenamefont {Ian},\ and\ \citenamefont {Sun}}]{chang2009}%
  \BibitemOpen
  \bibfield  {author} {\bibinfo {author} {\bibfnamefont {Y.}~\bibnamefont
  {Chang}}, \bibinfo {author} {\bibfnamefont {H.}~\bibnamefont {Ian}}, \ and\
  \bibinfo {author} {\bibfnamefont {C.}~\bibnamefont {Sun}},\ }\href@noop {}
  {\bibfield  {journal} {\bibinfo  {journal} {J. Phys. B}\ }\textbf {\bibinfo
  {volume} {42}},\ \bibinfo {pages} {215502} (\bibinfo {year}
  {2009})}\BibitemShut {NoStop}%
\bibitem [{\citenamefont {Wang}\ \emph {et~al.}(2010)\citenamefont {Wang},
  \citenamefont {Wang},\ and\ \citenamefont {Sun}}]{wang2010}%
  \BibitemOpen
  \bibfield  {author} {\bibinfo {author} {\bibfnamefont {W.}~\bibnamefont
  {Wang}}, \bibinfo {author} {\bibfnamefont {L.}~\bibnamefont {Wang}}, \ and\
  \bibinfo {author} {\bibfnamefont {H.}~\bibnamefont {Sun}},\ }\href@noop {}
  {\bibfield  {journal} {\bibinfo  {journal} {J. Korean Phys. Soc.}\ }\textbf
  {\bibinfo {volume} {57}},\ \bibinfo {pages} {704} (\bibinfo {year}
  {2010})}\BibitemShut {NoStop}%
\bibitem [{\citenamefont {Fisher}\ \emph {et~al.}(2016)\citenamefont {Fisher},
  \citenamefont {England}, \citenamefont {MacLean}, \citenamefont {Bustard},
  \citenamefont {Resch},\ and\ \citenamefont {Sussman}}]{fisher2016}%
  \BibitemOpen
  \bibfield  {author} {\bibinfo {author} {\bibfnamefont {K.~A.}\ \bibnamefont
  {Fisher}}, \bibinfo {author} {\bibfnamefont {D.~G.}\ \bibnamefont {England}},
  \bibinfo {author} {\bibfnamefont {J.-P.~W.}\ \bibnamefont {MacLean}},
  \bibinfo {author} {\bibfnamefont {P.~J.}\ \bibnamefont {Bustard}}, \bibinfo
  {author} {\bibfnamefont {K.~J.}\ \bibnamefont {Resch}}, \ and\ \bibinfo
  {author} {\bibfnamefont {B.~J.}\ \bibnamefont {Sussman}},\ }\href@noop {}
  {\bibfield  {journal} {\bibinfo  {journal} {Nat. Comm.}\ }\textbf {\bibinfo
  {volume} {7}} (\bibinfo {year} {2016})}\BibitemShut {NoStop}%
\bibitem [{\citenamefont {Lee}\ \emph {et~al.}(2012)\citenamefont {Lee},
  \citenamefont {Sussman}, \citenamefont {Sprague}, \citenamefont
  {Michelberger}, \citenamefont {Reim}, \citenamefont {Nunn}, \citenamefont
  {Langford}, \citenamefont {Bustard}, \citenamefont {Jaksch},\ and\
  \citenamefont {Walmsley}}]{lee2012}%
  \BibitemOpen
  \bibfield  {author} {\bibinfo {author} {\bibfnamefont {K.}~\bibnamefont
  {Lee}}, \bibinfo {author} {\bibfnamefont {B.}~\bibnamefont {Sussman}},
  \bibinfo {author} {\bibfnamefont {M.}~\bibnamefont {Sprague}}, \bibinfo
  {author} {\bibfnamefont {P.}~\bibnamefont {Michelberger}}, \bibinfo {author}
  {\bibfnamefont {K.}~\bibnamefont {Reim}}, \bibinfo {author} {\bibfnamefont
  {J.}~\bibnamefont {Nunn}}, \bibinfo {author} {\bibfnamefont {N.}~\bibnamefont
  {Langford}}, \bibinfo {author} {\bibfnamefont {P.}~\bibnamefont {Bustard}},
  \bibinfo {author} {\bibfnamefont {D.}~\bibnamefont {Jaksch}}, \ and\ \bibinfo
  {author} {\bibfnamefont {I.}~\bibnamefont {Walmsley}},\ }\href@noop {}
  {\bibfield  {journal} {\bibinfo  {journal} {Nat. Phot.}\ }\textbf {\bibinfo
  {volume} {6}},\ \bibinfo {pages} {41} (\bibinfo {year} {2012})}\BibitemShut
  {NoStop}%
\bibitem [{\citenamefont {Leibfried}\ \emph {et~al.}(2003)\citenamefont
  {Leibfried}, \citenamefont {Blatt}, \citenamefont {Monroe},\ and\
  \citenamefont {Wineland}}]{leibfried2003}%
  \BibitemOpen
  \bibfield  {author} {\bibinfo {author} {\bibfnamefont {D.}~\bibnamefont
  {Leibfried}}, \bibinfo {author} {\bibfnamefont {R.}~\bibnamefont {Blatt}},
  \bibinfo {author} {\bibfnamefont {C.}~\bibnamefont {Monroe}}, \ and\ \bibinfo
  {author} {\bibfnamefont {D.}~\bibnamefont {Wineland}},\ }\href@noop {}
  {\bibfield  {journal} {\bibinfo  {journal} {Rev. of Mod. Phys.}\ }\textbf
  {\bibinfo {volume} {75}},\ \bibinfo {pages} {281} (\bibinfo {year}
  {2003})}\BibitemShut {NoStop}%
\bibitem [{\citenamefont {Joannopoulos}\ \emph {et~al.}(2011)\citenamefont
  {Joannopoulos}, \citenamefont {Johnson}, \citenamefont {Winn},\ and\
  \citenamefont {Meade}}]{joannopoulos2011photonic}%
  \BibitemOpen
  \bibfield  {author} {\bibinfo {author} {\bibfnamefont {J.~D.}\ \bibnamefont
  {Joannopoulos}}, \bibinfo {author} {\bibfnamefont {S.~G.}\ \bibnamefont
  {Johnson}}, \bibinfo {author} {\bibfnamefont {J.~N.}\ \bibnamefont {Winn}}, \
  and\ \bibinfo {author} {\bibfnamefont {R.~D.}\ \bibnamefont {Meade}},\
  }\href@noop {} {\emph {\bibinfo {title} {Photonic crystals: molding the flow
  of light}}}\ (\bibinfo  {publisher} {Princeton university press},\ \bibinfo
  {year} {2011})\BibitemShut {NoStop}%
\bibitem [{Sup()}]{SuppInfo}%
  \BibitemOpen
  \href@noop {} {\bibinfo  {journal} {Supplemental Information}\ }\BibitemShut
  {NoStop}%
\bibitem [{\citenamefont {Para{\"\i}so}\ \emph {et~al.}(2015)\citenamefont
  {Para{\"\i}so}, \citenamefont {Kalaee}, \citenamefont {Zang}, \citenamefont
  {Pfeifer}, \citenamefont {Marquardt},\ and\ \citenamefont
  {Painter}}]{paraiso201}%
  \BibitemOpen
\bibfield  {journal} {  }\bibfield  {author} {\bibinfo {author} {\bibfnamefont
  {T.~K.}\ \bibnamefont {Para{\"\i}so}}, \bibinfo {author} {\bibfnamefont
  {M.}~\bibnamefont {Kalaee}}, \bibinfo {author} {\bibfnamefont
  {L.}~\bibnamefont {Zang}}, \bibinfo {author} {\bibfnamefont {H.}~\bibnamefont
  {Pfeifer}}, \bibinfo {author} {\bibfnamefont {F.}~\bibnamefont {Marquardt}},
  \ and\ \bibinfo {author} {\bibfnamefont {O.}~\bibnamefont {Painter}},\
  }\href@noop {} {\bibfield  {journal} {\bibinfo  {journal} {Phys. Rev. X}\
  }\textbf {\bibinfo {volume} {5}},\ \bibinfo {pages} {041024} (\bibinfo {year}
  {2015})}\BibitemShut {NoStop}%
\bibitem [{\citenamefont {Ludwig}\ \emph {et~al.}(2012)\citenamefont {Ludwig},
  \citenamefont {Safavi-Naeini}, \citenamefont {Painter},\ and\ \citenamefont
  {Marquardt}}]{ludwig2012}%
  \BibitemOpen
  \bibfield  {author} {\bibinfo {author} {\bibfnamefont {M.}~\bibnamefont
  {Ludwig}}, \bibinfo {author} {\bibfnamefont {A.~H.}\ \bibnamefont
  {Safavi-Naeini}}, \bibinfo {author} {\bibfnamefont {O.}~\bibnamefont
  {Painter}}, \ and\ \bibinfo {author} {\bibfnamefont {F.}~\bibnamefont
  {Marquardt}},\ }\href@noop {} {\bibfield  {journal} {\bibinfo  {journal}
  {Phys. Rev. Lett.}\ }\textbf {\bibinfo {volume} {109}},\ \bibinfo {pages}
  {063601} (\bibinfo {year} {2012})}\BibitemShut {NoStop}%
\bibitem [{\citenamefont {Johne}\ \emph {et~al.}(2015)\citenamefont {Johne},
  \citenamefont {Schutjens}, \citenamefont {Fattah~poor}, \citenamefont {Jin},\
  and\ \citenamefont {Fiore}}]{johne2015}%
  \BibitemOpen
  \bibfield  {author} {\bibinfo {author} {\bibfnamefont {R.}~\bibnamefont
  {Johne}}, \bibinfo {author} {\bibfnamefont {R.}~\bibnamefont {Schutjens}},
  \bibinfo {author} {\bibfnamefont {S.}~\bibnamefont {Fattah~poor}}, \bibinfo
  {author} {\bibfnamefont {C.-Y.}\ \bibnamefont {Jin}}, \ and\ \bibinfo
  {author} {\bibfnamefont {A.}~\bibnamefont {Fiore}},\ }\href@noop {}
  {\bibfield  {journal} {\bibinfo  {journal} {Phys. Rev. A}\ }\textbf {\bibinfo
  {volume} {91}},\ \bibinfo {pages} {063807} (\bibinfo {year}
  {2015})}\BibitemShut {NoStop}%
\bibitem [{\citenamefont {Caselli}\ \emph {et~al.}(2015)\citenamefont
  {Caselli}, \citenamefont {Riboli}, \citenamefont {La~China}, \citenamefont
  {Gerardino}, \citenamefont {Li}, \citenamefont {Linfield}, \citenamefont
  {Pagliano}, \citenamefont {Fiore}, \citenamefont {Intonti},\ and\
  \citenamefont {Gurioli}}]{caselli2015}%
  \BibitemOpen
  \bibfield  {author} {\bibinfo {author} {\bibfnamefont {N.}~\bibnamefont
  {Caselli}}, \bibinfo {author} {\bibfnamefont {F.}~\bibnamefont {Riboli}},
  \bibinfo {author} {\bibfnamefont {F.}~\bibnamefont {La~China}}, \bibinfo
  {author} {\bibfnamefont {A.}~\bibnamefont {Gerardino}}, \bibinfo {author}
  {\bibfnamefont {L.}~\bibnamefont {Li}}, \bibinfo {author} {\bibfnamefont
  {E.~H.}\ \bibnamefont {Linfield}}, \bibinfo {author} {\bibfnamefont
  {F.}~\bibnamefont {Pagliano}}, \bibinfo {author} {\bibfnamefont
  {A.}~\bibnamefont {Fiore}}, \bibinfo {author} {\bibfnamefont
  {F.}~\bibnamefont {Intonti}}, \ and\ \bibinfo {author} {\bibfnamefont
  {M.}~\bibnamefont {Gurioli}},\ }\href@noop {} {\bibfield  {journal} {\bibinfo
   {journal} {ACS Photonics}\ }\textbf {\bibinfo {volume} {2}},\ \bibinfo
  {pages} {565} (\bibinfo {year} {2015})}\BibitemShut {NoStop}%
\bibitem [{\citenamefont {Chan}\ \emph {et~al.}(2011)\citenamefont {Chan},
  \citenamefont {Alegre}, \citenamefont {Safavi-Naeini}, \citenamefont {Hill},
  \citenamefont {Krause}, \citenamefont {Gr{\"o}blacher}, \citenamefont
  {Aspelmeyer},\ and\ \citenamefont {Painter}}]{chan2011}%
  \BibitemOpen
  \bibfield  {author} {\bibinfo {author} {\bibfnamefont {J.}~\bibnamefont
  {Chan}}, \bibinfo {author} {\bibfnamefont {T.~M.}\ \bibnamefont {Alegre}},
  \bibinfo {author} {\bibfnamefont {A.~H.}\ \bibnamefont {Safavi-Naeini}},
  \bibinfo {author} {\bibfnamefont {J.~T.}\ \bibnamefont {Hill}}, \bibinfo
  {author} {\bibfnamefont {A.}~\bibnamefont {Krause}}, \bibinfo {author}
  {\bibfnamefont {S.}~\bibnamefont {Gr{\"o}blacher}}, \bibinfo {author}
  {\bibfnamefont {M.}~\bibnamefont {Aspelmeyer}}, \ and\ \bibinfo {author}
  {\bibfnamefont {O.}~\bibnamefont {Painter}},\ }\href@noop {} {\bibfield
  {journal} {\bibinfo  {journal} {Nature}\ }\textbf {\bibinfo {volume} {478}},\
  \bibinfo {pages} {89} (\bibinfo {year} {2011})}\BibitemShut {NoStop}%
\bibitem [{\citenamefont {Burek}\ \emph {et~al.}(2015)\citenamefont {Burek},
  \citenamefont {Cohen}, \citenamefont {Meenehan}, \citenamefont {Ruelle},
  \citenamefont {Meesala}, \citenamefont {Rochman}, \citenamefont {Atikian},
  \citenamefont {Markham}, \citenamefont {Twitchen}, \citenamefont {Lukin},
  \citenamefont {Painter},\ and\ \citenamefont {Lon{\v
  c}ar}}]{burek2015diamond}%
  \BibitemOpen
  \bibfield  {author} {\bibinfo {author} {\bibfnamefont {M.~J.}\ \bibnamefont
  {Burek}}, \bibinfo {author} {\bibfnamefont {J.~D.}\ \bibnamefont {Cohen}},
  \bibinfo {author} {\bibfnamefont {S.~M.}\ \bibnamefont {Meenehan}}, \bibinfo
  {author} {\bibfnamefont {T.}~\bibnamefont {Ruelle}}, \bibinfo {author}
  {\bibfnamefont {S.}~\bibnamefont {Meesala}}, \bibinfo {author} {\bibfnamefont
  {J.}~\bibnamefont {Rochman}}, \bibinfo {author} {\bibfnamefont {H.~A.}\
  \bibnamefont {Atikian}}, \bibinfo {author} {\bibfnamefont {M.}~\bibnamefont
  {Markham}}, \bibinfo {author} {\bibfnamefont {D.~J.}\ \bibnamefont
  {Twitchen}}, \bibinfo {author} {\bibfnamefont {M.~D.}\ \bibnamefont {Lukin}},
  \bibinfo {author} {\bibfnamefont {O.}~\bibnamefont {Painter}}, \ and\
  \bibinfo {author} {\bibfnamefont {M.}~\bibnamefont {Lon{\v c}ar}},\
  }\href@noop {} {\bibfield  {journal} {\bibinfo  {journal} {arXiv:1512.04166}\
  } (\bibinfo {year} {2015})}\BibitemShut {NoStop}%
\bibitem [{\citenamefont {Johansson}\ \emph {et~al.}(2012)\citenamefont
  {Johansson}, \citenamefont {Nation},\ and\ \citenamefont
  {Nori}}]{johansson2012qutip}%
  \BibitemOpen
  \bibfield  {author} {\bibinfo {author} {\bibfnamefont {J.}~\bibnamefont
  {Johansson}}, \bibinfo {author} {\bibfnamefont {P.}~\bibnamefont {Nation}}, \
  and\ \bibinfo {author} {\bibfnamefont {F.}~\bibnamefont {Nori}},\ }\href@noop
  {} {\bibfield  {journal} {\bibinfo  {journal} {Comput. Phys. Commun.}\
  }\textbf {\bibinfo {volume} {183}},\ \bibinfo {pages} {1760} (\bibinfo {year}
  {2012})}\BibitemShut {NoStop}%
\bibitem [{\citenamefont {Lee}\ \emph {et~al.}(2014)\citenamefont {Lee},
  \citenamefont {Bracher}, \citenamefont {Cui}, \citenamefont {Ohno},
  \citenamefont {McLellan}, \citenamefont {Zhang}, \citenamefont {Andrich},
  \citenamefont {Alem{\'a}n}, \citenamefont {Russell}, \citenamefont {Magyar},
  \citenamefont {Aharonovich}, \citenamefont {Jayich}, \citenamefont
  {Awschalom},\ and\ \citenamefont {Hu}}]{lee2014}%
  \BibitemOpen
  \bibfield  {author} {\bibinfo {author} {\bibfnamefont {J.~C.}\ \bibnamefont
  {Lee}}, \bibinfo {author} {\bibfnamefont {D.~O.}\ \bibnamefont {Bracher}},
  \bibinfo {author} {\bibfnamefont {S.}~\bibnamefont {Cui}}, \bibinfo {author}
  {\bibfnamefont {K.}~\bibnamefont {Ohno}}, \bibinfo {author} {\bibfnamefont
  {C.~A.}\ \bibnamefont {McLellan}}, \bibinfo {author} {\bibfnamefont
  {X.}~\bibnamefont {Zhang}}, \bibinfo {author} {\bibfnamefont
  {P.}~\bibnamefont {Andrich}}, \bibinfo {author} {\bibfnamefont
  {B.}~\bibnamefont {Alem{\'a}n}}, \bibinfo {author} {\bibfnamefont {K.~J.}\
  \bibnamefont {Russell}}, \bibinfo {author} {\bibfnamefont {A.~P.}\
  \bibnamefont {Magyar}}, \bibinfo {author} {\bibfnamefont {I.}~\bibnamefont
  {Aharonovich}}, \bibinfo {author} {\bibfnamefont {A.~B.}\ \bibnamefont
  {Jayich}}, \bibinfo {author} {\bibfnamefont {D.}~\bibnamefont {Awschalom}}, \
  and\ \bibinfo {author} {\bibfnamefont {E.~L.}\ \bibnamefont {Hu}},\
  }\href@noop {} {\bibfield  {journal} {\bibinfo  {journal} {Appl. Phys.
  Lett.}\ }\textbf {\bibinfo {volume} {105}},\ \bibinfo {pages} {261101}
  (\bibinfo {year} {2014})}\BibitemShut {NoStop}%
\bibitem [{\citenamefont {Goban}\ \emph {et~al.}(2014)\citenamefont {Goban},
  \citenamefont {Hung}, \citenamefont {Yu}, \citenamefont {Hood}, \citenamefont
  {Muniz}, \citenamefont {Lee}, \citenamefont {Martin}, \citenamefont
  {McClung}, \citenamefont {Choi}, \citenamefont {Chang}, \citenamefont
  {Painter},\ and\ \citenamefont {Kimble}}]{goban2014atom}%
  \BibitemOpen
  \bibfield  {author} {\bibinfo {author} {\bibfnamefont {A.}~\bibnamefont
  {Goban}}, \bibinfo {author} {\bibfnamefont {C.-L.}\ \bibnamefont {Hung}},
  \bibinfo {author} {\bibfnamefont {S.-P.}\ \bibnamefont {Yu}}, \bibinfo
  {author} {\bibfnamefont {J.}~\bibnamefont {Hood}}, \bibinfo {author}
  {\bibfnamefont {J.}~\bibnamefont {Muniz}}, \bibinfo {author} {\bibfnamefont
  {J.}~\bibnamefont {Lee}}, \bibinfo {author} {\bibfnamefont {M.}~\bibnamefont
  {Martin}}, \bibinfo {author} {\bibfnamefont {A.}~\bibnamefont {McClung}},
  \bibinfo {author} {\bibfnamefont {K.}~\bibnamefont {Choi}}, \bibinfo {author}
  {\bibfnamefont {D.}~\bibnamefont {Chang}}, \bibinfo {author} {\bibfnamefont
  {O.}~\bibnamefont {Painter}}, \ and\ \bibinfo {author} {\bibfnamefont
  {H.}~\bibnamefont {Kimble}},\ }\href@noop {} {\bibfield  {journal} {\bibinfo
  {journal} {Nat. Comm.}\ }\textbf {\bibinfo {volume} {5}} (\bibinfo {year}
  {2014})}\BibitemShut {NoStop}%
\bibitem [{\citenamefont {Thompson}\ \emph {et~al.}(2013)\citenamefont
  {Thompson}, \citenamefont {Tiecke}, \citenamefont {de~Leon}, \citenamefont
  {Feist}, \citenamefont {Akimov}, \citenamefont {Gullans}, \citenamefont
  {Zibrov}, \citenamefont {Vuleti{\'c}},\ and\ \citenamefont
  {Lukin}}]{thompson2013coupling}%
  \BibitemOpen
  \bibfield  {author} {\bibinfo {author} {\bibfnamefont {J.}~\bibnamefont
  {Thompson}}, \bibinfo {author} {\bibfnamefont {T.}~\bibnamefont {Tiecke}},
  \bibinfo {author} {\bibfnamefont {N.}~\bibnamefont {de~Leon}}, \bibinfo
  {author} {\bibfnamefont {J.}~\bibnamefont {Feist}}, \bibinfo {author}
  {\bibfnamefont {A.}~\bibnamefont {Akimov}}, \bibinfo {author} {\bibfnamefont
  {M.}~\bibnamefont {Gullans}}, \bibinfo {author} {\bibfnamefont
  {A.}~\bibnamefont {Zibrov}}, \bibinfo {author} {\bibfnamefont
  {V.}~\bibnamefont {Vuleti{\'c}}}, \ and\ \bibinfo {author} {\bibfnamefont
  {M.}~\bibnamefont {Lukin}},\ }\href@noop {} {\bibfield  {journal} {\bibinfo
  {journal} {Science}\ }\textbf {\bibinfo {volume} {340}},\ \bibinfo {pages}
  {1202} (\bibinfo {year} {2013})}\BibitemShut {NoStop}%
\bibitem [{\citenamefont {Hammerer}\ \emph {et~al.}(2009)\citenamefont
  {Hammerer}, \citenamefont {Wallquist}, \citenamefont {Genes}, \citenamefont
  {Ludwig}, \citenamefont {Marquardt}, \citenamefont {Treutlein}, \citenamefont
  {Zoller}, \citenamefont {Ye},\ and\ \citenamefont
  {Kimble}}]{hammerer2009strong}%
  \BibitemOpen
  \bibfield  {author} {\bibinfo {author} {\bibfnamefont {K.}~\bibnamefont
  {Hammerer}}, \bibinfo {author} {\bibfnamefont {M.}~\bibnamefont {Wallquist}},
  \bibinfo {author} {\bibfnamefont {C.}~\bibnamefont {Genes}}, \bibinfo
  {author} {\bibfnamefont {M.}~\bibnamefont {Ludwig}}, \bibinfo {author}
  {\bibfnamefont {F.}~\bibnamefont {Marquardt}}, \bibinfo {author}
  {\bibfnamefont {P.}~\bibnamefont {Treutlein}}, \bibinfo {author}
  {\bibfnamefont {P.}~\bibnamefont {Zoller}}, \bibinfo {author} {\bibfnamefont
  {J.}~\bibnamefont {Ye}}, \ and\ \bibinfo {author} {\bibfnamefont {H.~J.}\
  \bibnamefont {Kimble}},\ }\href@noop {} {\bibfield  {journal} {\bibinfo
  {journal} {Phys. Rev. Lett.}\ }\textbf {\bibinfo {volume} {103}},\ \bibinfo
  {pages} {063005} (\bibinfo {year} {2009})}\BibitemShut {NoStop}%
\end{thebibliography}

\begin{thebibliography}{10}%
\makeatletter
\providecommand \@ifxundefined [1]{%
 \@ifx{#1\undefined}
}%
\providecommand \@ifnum [1]{%
 \ifnum #1\expandafter \@firstoftwo
 \else \expandafter \@secondoftwo
 \fi
}%
\providecommand \@ifx [1]{%
 \ifx #1\expandafter \@firstoftwo
 \else \expandafter \@secondoftwo
 \fi
}%
\providecommand \natexlab [1]{#1}%
\providecommand \enquote  [1]{``#1''}%
\providecommand \bibnamefont  [1]{#1}%
\providecommand \bibfnamefont [1]{#1}%
\providecommand \citenamefont [1]{#1}%
\providecommand \href@noop [0]{\@secondoftwo}%
\providecommand \href [0]{\begingroup \@sanitize@url \@href}%
\providecommand \@href[1]{\@@startlink{#1}\@@href}%
\providecommand \@@href[1]{\endgroup#1\@@endlink}%
\providecommand \@sanitize@url [0]{\catcode `\\12\catcode `\$12\catcode
  `\&12\catcode `\#12\catcode `\^12\catcode `\_12\catcode `\%12\relax}%
\providecommand \@@startlink[1]{}%
\providecommand \@@endlink[0]{}%
\providecommand \url  [0]{\begingroup\@sanitize@url \@url }%
\providecommand \@url [1]{\endgroup\@href {#1}{\urlprefix }}%
\providecommand \urlprefix  [0]{URL }%
\providecommand \Eprint [0]{\href }%
\providecommand \doibase [0]{http://dx.doi.org/}%
\providecommand \selectlanguage [0]{\@gobble}%
\providecommand \bibinfo  [0]{\@secondoftwo}%
\providecommand \bibfield  [0]{\@secondoftwo}%
\providecommand \translation [1]{[#1]}%
\providecommand \BibitemOpen [0]{}%
\providecommand \bibitemStop [0]{}%
\providecommand \bibitemNoStop [0]{.\EOS\space}%
\providecommand \EOS [0]{\spacefactor3000\relax}%
\providecommand \BibitemShut  [1]{\csname bibitem#1\endcsname}%
\let\auto@bib@innerbib\@empty
\bibitem [{\citenamefont {Johansson}\ \emph {et~al.}(2012)\citenamefont
  {Johansson}, \citenamefont {Nation},\ and\ \citenamefont
  {Nori}}]{SI_johansson2012qutip}%
  \BibitemOpen
  \bibfield  {author} {\bibinfo {author} {\bibfnamefont {J.}~\bibnamefont
  {Johansson}}, \bibinfo {author} {\bibfnamefont {P.}~\bibnamefont {Nation}}, \
  and\ \bibinfo {author} {\bibfnamefont {F.}~\bibnamefont {Nori}},\ }\href@noop
  {} {\bibfield  {journal} {\bibinfo  {journal} {Comput. Phys. Commun.}\
  }\textbf {\bibinfo {volume} {183}},\ \bibinfo {pages} {1760} (\bibinfo {year}
  {2012})}\BibitemShut {NoStop}%
\bibitem [{\citenamefont {Ludwig}\ \emph {et~al.}(2012)\citenamefont {Ludwig},
  \citenamefont {Safavi-Naeini}, \citenamefont {Painter},\ and\ \citenamefont
  {Marquardt}}]{SI_ludwig2012}%
  \BibitemOpen
  \bibfield  {author} {\bibinfo {author} {\bibfnamefont {M.}~\bibnamefont
  {Ludwig}}, \bibinfo {author} {\bibfnamefont {A.~H.}\ \bibnamefont
  {Safavi-Naeini}}, \bibinfo {author} {\bibfnamefont {O.}~\bibnamefont
  {Painter}}, \ and\ \bibinfo {author} {\bibfnamefont {F.}~\bibnamefont
  {Marquardt}},\ }\href@noop {} {\bibfield  {journal} {\bibinfo  {journal}
  {Phys. Rev. Lett.}\ }\textbf {\bibinfo {volume} {109}},\ \bibinfo {pages}
  {063601} (\bibinfo {year} {2012})}\BibitemShut {NoStop}%
\bibitem [{\citenamefont {Lee}\ \emph {et~al.}(2014)\citenamefont {Lee},
  \citenamefont {Bracher}, \citenamefont {Cui}, \citenamefont {Ohno},
  \citenamefont {McLellan}, \citenamefont {Zhang}, \citenamefont {Andrich},
  \citenamefont {Alem{\'a}n}, \citenamefont {Russell}, \citenamefont {Magyar},
  \citenamefont {Aharonovich}, \citenamefont {Jayich}, \citenamefont
  {Awschalom},\ and\ \citenamefont {Hu}}]{SI_lee2014}%
  \BibitemOpen
  \bibfield  {author} {\bibinfo {author} {\bibfnamefont {J.~C.}\ \bibnamefont
  {Lee}}, \bibinfo {author} {\bibfnamefont {D.~O.}\ \bibnamefont {Bracher}},
  \bibinfo {author} {\bibfnamefont {S.}~\bibnamefont {Cui}}, \bibinfo {author}
  {\bibfnamefont {K.}~\bibnamefont {Ohno}}, \bibinfo {author} {\bibfnamefont
  {C.~A.}\ \bibnamefont {McLellan}}, \bibinfo {author} {\bibfnamefont
  {X.}~\bibnamefont {Zhang}}, \bibinfo {author} {\bibfnamefont
  {P.}~\bibnamefont {Andrich}}, \bibinfo {author} {\bibfnamefont
  {B.}~\bibnamefont {Alem{\'a}n}}, \bibinfo {author} {\bibfnamefont {K.~J.}\
  \bibnamefont {Russell}}, \bibinfo {author} {\bibfnamefont {A.~P.}\
  \bibnamefont {Magyar}}, \bibinfo {author} {\bibfnamefont {I.}~\bibnamefont
  {Aharonovich}}, \bibinfo {author} {\bibfnamefont {A.~B.}\ \bibnamefont
  {Jayich}}, \bibinfo {author} {\bibfnamefont {D.}~\bibnamefont {Awschalom}}, \
  and\ \bibinfo {author} {\bibfnamefont {E.~L.}\ \bibnamefont {Hu}},\
  }\href@noop {} {\bibfield  {journal} {\bibinfo  {journal} {Appl. Phys.
  Lett.}\ }\textbf {\bibinfo {volume} {105}},\ \bibinfo {pages} {261101}
  (\bibinfo {year} {2014})}\BibitemShut {NoStop}%
\bibitem [{\citenamefont {Faraon}\ \emph {et~al.}(2012)\citenamefont {Faraon},
  \citenamefont {Santori}, \citenamefont {Huang}, \citenamefont {Acosta},\ and\
  \citenamefont {Beausoleil}}]{SI_faraon2012coupling}%
  \BibitemOpen
  \bibfield  {author} {\bibinfo {author} {\bibfnamefont {A.}~\bibnamefont
  {Faraon}}, \bibinfo {author} {\bibfnamefont {C.}~\bibnamefont {Santori}},
  \bibinfo {author} {\bibfnamefont {Z.}~\bibnamefont {Huang}}, \bibinfo
  {author} {\bibfnamefont {V.~M.}\ \bibnamefont {Acosta}}, \ and\ \bibinfo
  {author} {\bibfnamefont {R.~G.}\ \bibnamefont {Beausoleil}},\ }\href@noop {}
  {\bibfield  {journal} {\bibinfo  {journal} {Phys. Rev. Lett.}\ }\textbf
  {\bibinfo {volume} {109}},\ \bibinfo {pages} {033604} (\bibinfo {year}
  {2012})}\BibitemShut {NoStop}%
\bibitem [{\citenamefont {Lodahl}\ \emph {et~al.}(2015)\citenamefont {Lodahl},
  \citenamefont {Mahmoodian},\ and\ \citenamefont
  {Stobbe}}]{SI_lodahl2015interfacing}%
  \BibitemOpen
  \bibfield  {author} {\bibinfo {author} {\bibfnamefont {P.}~\bibnamefont
  {Lodahl}}, \bibinfo {author} {\bibfnamefont {S.}~\bibnamefont {Mahmoodian}},
  \ and\ \bibinfo {author} {\bibfnamefont {S.}~\bibnamefont {Stobbe}},\
  }\href@noop {} {\bibfield  {journal} {\bibinfo  {journal} {Rev. Mod. Phys.}\
  }\textbf {\bibinfo {volume} {87}},\ \bibinfo {pages} {347} (\bibinfo {year}
  {2015})}\BibitemShut {NoStop}%
\bibitem [{\citenamefont {Robledo}\ \emph {et~al.}(2010)\citenamefont
  {Robledo}, \citenamefont {Bernien}, \citenamefont {van Weperen},\ and\
  \citenamefont {Hanson}}]{SI_robledo2010control}%
  \BibitemOpen
  \bibfield  {author} {\bibinfo {author} {\bibfnamefont {L.}~\bibnamefont
  {Robledo}}, \bibinfo {author} {\bibfnamefont {H.}~\bibnamefont {Bernien}},
  \bibinfo {author} {\bibfnamefont {I.}~\bibnamefont {van Weperen}}, \ and\
  \bibinfo {author} {\bibfnamefont {R.}~\bibnamefont {Hanson}},\ }\href@noop {}
  {\bibfield  {journal} {\bibinfo  {journal} {Phys. Rev. Lett.}\ }\textbf
  {\bibinfo {volume} {105}},\ \bibinfo {pages} {177403} (\bibinfo {year}
  {2010})}\BibitemShut {NoStop}%
\bibitem [{\citenamefont {Burek}\ \emph {et~al.}(2015)\citenamefont {Burek},
  \citenamefont {Cohen}, \citenamefont {Meenehan}, \citenamefont {Ruelle},
  \citenamefont {Meesala}, \citenamefont {Rochman}, \citenamefont {Atikian},
  \citenamefont {Markham}, \citenamefont {Twitchen}, \citenamefont {Lukin},
  \citenamefont {Painter},\ and\ \citenamefont {Lon{\v
  c}ar}}]{SI_burek2015diamond}%
  \BibitemOpen
  \bibfield  {author} {\bibinfo {author} {\bibfnamefont {M.~J.}\ \bibnamefont
  {Burek}}, \bibinfo {author} {\bibfnamefont {J.~D.}\ \bibnamefont {Cohen}},
  \bibinfo {author} {\bibfnamefont {S.~M.}\ \bibnamefont {Meenehan}}, \bibinfo
  {author} {\bibfnamefont {T.}~\bibnamefont {Ruelle}}, \bibinfo {author}
  {\bibfnamefont {S.}~\bibnamefont {Meesala}}, \bibinfo {author} {\bibfnamefont
  {J.}~\bibnamefont {Rochman}}, \bibinfo {author} {\bibfnamefont {H.~A.}\
  \bibnamefont {Atikian}}, \bibinfo {author} {\bibfnamefont {M.}~\bibnamefont
  {Markham}}, \bibinfo {author} {\bibfnamefont {D.~J.}\ \bibnamefont
  {Twitchen}}, \bibinfo {author} {\bibfnamefont {M.~D.}\ \bibnamefont {Lukin}},
  \bibinfo {author} {\bibfnamefont {O.}~\bibnamefont {Painter}}, \ and\
  \bibinfo {author} {\bibfnamefont {M.}~\bibnamefont {Lon{\v c}ar}},\
  }\href@noop {} {\bibfield  {journal} {\bibinfo  {journal} {arXiv:1512.04166}\
  } (\bibinfo {year} {2015})}\BibitemShut {NoStop}%
\bibitem [{\citenamefont {Barzanjeh}\ \emph {et~al.}(2011)\citenamefont
  {Barzanjeh}, \citenamefont {Naderi},\ and\ \citenamefont
  {Soltanolkotabi}}]{SI_barzanjeh2011}%
  \BibitemOpen
  \bibfield  {author} {\bibinfo {author} {\bibfnamefont {S.}~\bibnamefont
  {Barzanjeh}}, \bibinfo {author} {\bibfnamefont {M.~H.}\ \bibnamefont
  {Naderi}}, \ and\ \bibinfo {author} {\bibfnamefont {M.}~\bibnamefont
  {Soltanolkotabi}},\ }\href@noop {} {\bibfield  {journal} {\bibinfo  {journal}
  {Phys. Rev. A}\ }\textbf {\bibinfo {volume} {84}},\ \bibinfo {pages} {063850}
  (\bibinfo {year} {2011})}\BibitemShut {NoStop}%
\bibitem [{\citenamefont {Chang}\ \emph {et~al.}(2009)\citenamefont {Chang},
  \citenamefont {Ian},\ and\ \citenamefont {Sun}}]{SI_chang2009}%
  \BibitemOpen
  \bibfield  {author} {\bibinfo {author} {\bibfnamefont {Y.}~\bibnamefont
  {Chang}}, \bibinfo {author} {\bibfnamefont {H.}~\bibnamefont {Ian}}, \ and\
  \bibinfo {author} {\bibfnamefont {C.}~\bibnamefont {Sun}},\ }\href@noop {}
  {\bibfield  {journal} {\bibinfo  {journal} {J. Phys. B}\ }\textbf {\bibinfo
  {volume} {42}},\ \bibinfo {pages} {215502} (\bibinfo {year}
  {2009})}\BibitemShut {NoStop}%
\bibitem [{\citenamefont {Wang}\ \emph {et~al.}(2010)\citenamefont {Wang},
  \citenamefont {Wang},\ and\ \citenamefont {Sun}}]{SI_wang2010}%
  \BibitemOpen
  \bibfield  {author} {\bibinfo {author} {\bibfnamefont {W.}~\bibnamefont
  {Wang}}, \bibinfo {author} {\bibfnamefont {L.}~\bibnamefont {Wang}}, \ and\
  \bibinfo {author} {\bibfnamefont {H.}~\bibnamefont {Sun}},\ }\href@noop {}
  {\bibfield  {journal} {\bibinfo  {journal} {J. Korean Phys. Soc.}\ }\textbf
  {\bibinfo {volume} {57}},\ \bibinfo {pages} {704} (\bibinfo {year}
  {2010})}\BibitemShut {NoStop}%
\end{thebibliography}
\end{document}